\documentclass[12pt]{article}
\usepackage[margin=0.75in]{geometry}
\usepackage{authblk}

\usepackage{amsfonts}
\usepackage{amsmath}
\usepackage{amssymb}
\usepackage{physics}
\usepackage{braket}
\usepackage{slashed}
\usepackage{mathtools}
\usepackage{url}
\usepackage{breqn}
\usepackage{color}
\usepackage{array}
\usepackage{graphicx}
\graphicspath{{figs/}}
\usepackage{bbm}
\usepackage{subcaption}
\usepackage{float}
\usepackage{bm}
\usepackage{upgreek}
\usepackage[labelfont=bf, font = small]{caption}
\usepackage{enumitem}				      
\usepackage[textsize=tiny]{todonotes}

\usepackage[linktocpage]{hyperref}
\hypersetup{
    colorlinks = true,
    urlcolor = magenta,
    linkcolor = blue,
    citecolor = magenta,
  }
\usepackage[sorting=none, style=numeric-comp,giveninits,maxbibnames=4,maxcitenames=4]{biblatex}
\renewbibmacro{in:}{}
\DeclareFieldFormat{pages}{#1}
\DeclareFieldFormat[article]{title}{}
\DeclareFieldFormat[article]{journaltitle}{#1}
\DeclareFieldFormat[inproceedings]{title}{}
\DeclareFieldFormat[inproceedings]{booktitle}{In \textit{#1}}
\DeclareFieldFormat[inproceedings]{pages}{pp. #1}

\usepackage{siunitx}
\usepackage{amsthm}
\usepackage[textsize=tiny]{todonotes}

\usepackage{booktabs}

\newcommand{\realp}{\Re {\rm e}}
\newcommand{\Z}{\mathbb{Z}}

\newcommand{\diag}{\text{diag}}
\newcommand{\tot}{\text{tot}}

\newcommand{\pl}{\text{Pl}}
\newcommand{\mmin}{|\tilde{m}_{\text{min}}|}
\DeclareSIUnit\parsec{pc}

\newcommand{\mpl}{m_{\pl}}

\newcommand{\gw}{{\text{gw}}}
\newcommand{\const}{{\text{const}}}

\DeclareMathOperator\arcosh{arcosh}

\usepackage{dsfont}
\newcommand{\id}{\mathds{1}}

\numberwithin{equation}{section}

\title{Inflaton Self Resonance, Oscillons, and Gravitational Waves in Small Field Polynomial Inflation}
\author{Manuel Drees\footnote{drees@th.physik.uni-bonn.de}}
\author{Chenhuan Wang\footnote{cwang1@uni-bonn.de} }
\affil{Bethe Center for Theoretical Physics and Physikalisches Institut, Universit{\"a}t Bonn, \\
Nussallee 12, 53115 Bonn, Germany}

\date{\today}

\begin{document}
\maketitle
\begin{abstract}
  In this work, we investigate the post--inflationary dynamics of a
  simple single--field model with a renormalizable inflaton potential
  featuring a near--inflection point at a field value $\phi_0$. Due to
  the concave shape of the scalar potential, the effective mass of the
  inflaton becomes imaginary during as well as for some period after
  slow--roll inflation. As a result, in the initial reheating phase,
  where the inflaton oscillates around its minimum with a large
  amplitude, some field fluctuations grow exponentially; this effect
  becomes stronger at smaller $\phi_0$. This aspect can be analyzed
  using the Floquet theorem. We also analytically estimate the back
  reaction time after which the perturbations affect the evolution of
  the average inflaton field. In order to fully analyze this
  non--perturbative regime, we perform a (classical) lattice
  simulation, which reveals that the exponential growth of field
  fluctuations can fragment the system. This leads to a large amount
  of non--Gaussianity at very small scales, but the equation of state
  remains close to matter--like. The evolution of the background field
  throughout the fragmentation phase can be understood using the
  Hartree approximation. For sufficiently small $\phi_0$ soliton--like
  objects, called oscillons in the literature, are formed. This leads
  to areas with high local over--density, $\delta \rho \gg \bar\rho$
  where $\bar\rho$ is the average energy density. We speculate that
  this could lead to the formation of light primordial black holes,
  with lifetime $\gtrsim 10^{-19} \si{\sec}$. Other possibly
  observational consequences, in particular gravitational waves in the
  $\si{\mega\hertz}-\si{\giga\hertz}$ range, are discussed as
  well. Although a complete analytical study is difficult in our case,
  we obtain a power law scaling for the potential observables on
  $\phi_0$.
\end{abstract}

\section{Introduction}
\label{sec:intro}

Inflation as a theory of the very early Universe predicts the origin of
density perturbations and their {\em coherent} phases after they
re--enter the horizon \cite{Baumann:2009ds, Dodelson:2003ip}. The
simplest realization, using a single inflaton field undergoing
slow--roll, remains compatible with all observations \cite{planck6} as
long as the scalar potential $V(\phi)$ is sufficiently flat. Moreover,
if inflation occurs for field values well below the Planck scale, a
spectral index of density perturbations below $1$, as indicated by
observation, can only be realized if the second derivative of the
potential is negative during inflation, leading to a {\em tachyonic}
inflaton mass. In many cases this tachyonic mass will persist at the
onset of reheating, when the inflaton field begins to oscillate around
its minimum.

It has been realized some time ago that there are non--perturbative
channels for the inflaton to transfer its energy to other fields, see
e.g. \cite{kofmanTheoryReheatingInflation1997, chungProbingPlanckianPhysics2000, 
    Copeland:2002ku, dufauxPreheatingTrilinearInteractions2006}. 
  Here we investigate the closely related tachyonic
resonance, which occurs due to the inflaton self--couplings. We
perform our analysis for a particularly simple model
\cite{dreesSmallFieldPolynomial2021}, which assumes a renormalizable
potential featuring a near--inflection point at a field value
$\phi_0$, with inflation occurring at $\phi$ just below $\phi_0$. We
find that the tachyonic resonance can lead to ample ``particle
production'', i.e. the production of modes of the inflaton field with
finite momentum.

In the model we consider, a tachyonic inflaton mass at large field
values implies that the potential is shallower than quadratic near the
minimum for $\phi > 0$. It is known that a shallower than quadratic
potential can lead \cite{aminFlattopOscillonsExpanding2010} to the formation of
soliton--like objects, called oscillons. We show that such objects are
also generated in our model, even though the potential is steeper than
quadratic for $\phi < 0$, i.e. for ``half'' of each oscillation of the
inflaton field. These oscillons are relatively long--lived and highly
localized. As a result, primordial black holes (PBHs) might be
produced \cite{lozanovGravitationalPerturbationsOscillons2019, kouOscillonPreheatingFull2020,
  nazariOscillonCollapseBlack2021}. 
  Moreover, during their formation the oscillons emit
high frequency gravitational waves (GWs)
\cite{huangArtLatticeGravity2011, zhouGravitationalWavesOscillon2013, 
    Antusch:2016con, antuschWhatCanWe2018, Amin:2018xfe, kouOscillonPreheatingFull2020, 
    lozanovEnhancedGravitationalWaves2022}. (See
also the review \cite{Amin:2014eta}.)

The remainder of this article is organized as follows. In section
\ref{sec:model}, the model we consider is briefly introduced and a
first hint for a self--resonance is presented. In section
\ref{sec:floquet}, we use the Floquet theorem in order to check the
amount and range of instability. For the final, credible result we use
lattice simulations; they are discussed in section
\ref{sec:lattice}. Possible observational consequences are analyzed in
section \ref{sec:obs}. Finally, the results will be summarized in
section \ref{sec:conc}.

In this work, we use the reduced Planck mass
$m_{\pl} = \SI{2.435e18}{\giga\eV}$. Derivatives with respect to the
cosmic time $t$ are denoted with a dot, e.g.
$\dot{\phi} = \partial \phi / \partial t$. The derivative of the
scalar potential is defined as
\begin{equation} \label{eq:deriv}
V'(\phi) = V_{,\phi} (\phi) = \partial_{\phi} V(\phi)\,.
\end{equation}
Similarly, $V''$ denotes the second derivative of $V(\phi)$. Except
where explicitly noted, $k$ refers to the comoving momentum or wave
vector, i.e.  $k = q a$ with $q$ the physical momentum and $a$ the
(dimensionless) scale factor; the latter is normalized such that
$a = 1$ at the end of inflation. The following tools
and programs are used in this project: {\tt SciPy 1.0} \cite{2020SciPy-NMeth},
{\tt NumPy} \cite{harris2020array}, {\tt Scikit-learn}
\cite{scikit-learn}, {\tt Matplotlib} \cite{Hunter:2007}, {\tt
  PyVista} \cite{sullivan2019pyvista}, {\tt HDF5}
\cite{collette_python_hdf5_2014}, {\tt CosmoLattice}
\cite{figueroaCosmoLattice2021}, and {\tt HLATTICE}
\cite{huangArtLatticeGravity2011}.

\section{Inflationary model}
\label{sec:model}

Our analysis is based on the model proposed in
\cite{dreesSmallFieldPolynomial2021}. It is the unique single--field
model with a purely renormalizable potential that agrees with
observations. The inflaton potential is given by
\begin{equation} \label{math:V-sfpi}
  V(\phi) = d \left[ \phi^2 - \frac{8}{3} (1 - \beta) \phi_0 \phi + 2 \phi_0^2
  \right] \phi^2\, .
\end{equation}
The observational upper bound on primordial tensor modes implies
$\phi_0 < 22 \ \mpl$ \cite{Drees:2022aea}. However, we will see
shortly that interesting nonlinear dynamics occurs in the
post--inflationary epoch only in the small--field version of the
model, where $\phi_0$ is well below $\mpl$. In this model, density
perturbations observed in the CMB were generated at field values just
below $\phi_0$. By fitting the Planck 2018 data
\cite{Planck2018Inflation}, in particular the central values of the
scalar power spectrum amplitude $A_s$ and of the scalar spectral index
$n_s$, and requiring $N_{\rm CMB} = 65$ e--folds of inflation after
the CMB pivot scale first crossed out of the horizon, the model
parameters are determined to be
\begin{equation} \label{par-values}
  \beta = \num{9.73e-7} \left( \frac{\phi_0}{\mpl} \right) ^4\,, \quad
  d = \num{6.61e-16} \left( \frac{\phi_0}{\mpl} \right) ^2\,;
\end{equation}
these numbers depend only weakly on $N_{\rm CMB}$
\cite{dreesSmallFieldPolynomial2021}. Thus, $\phi_0$ is essentially
the only free parameter left.

In order to recover the hot big bang, a reheating phase must be
invoked. To that end one can introduce a Yukawa coupling to fermions
and/or a cubic coupling $\phi |\phi'|^2$ to complex scalar fields
$\phi'$. In order to preserve the flatness of the potential necessary
for slow--roll inflation, one should require that the ($1-$loop)
radiative corrections due to these new couplings to the derivatives of
the inflaton potential don't exceed the tree level contributions. In
combination with the lower bound on the reheat temperature for
successful big bang nucleosynthesis, this leads to a lower bound on
$\phi_0$, $\phi_0 > \num{3.40e-5}\, \mpl$
\cite{dreesSmallFieldPolynomial2021}.

At least within the usual cosmological models, the Planck data require
the spectral index of the density perturbations to be smaller than $1$
\cite{Planck2018Inflation} at
\begin{equation} \label{ns-exp}
  n_s = 0.9659 \pm 0.0040\,.
\end{equation}
In slow--roll inflation, $n_s$ is predicted to be \cite{Lyth:2009zz}
\begin{equation} \label{ns-th}
  n_s = 1 - 6 \epsilon_V + 2 \eta_V\,,
\end{equation}
where the (potential) slow--roll parameters are given by
\begin{equation} \label{slow-roll-pars}
  \epsilon_V = \frac{\mpl^2}{2} \left( \frac {V'} {V} \right)^2\,; \quad
  \eta_V = \mpl^2 \frac { V''} {V}\,.
\end{equation}
In small--field inflation, one has generically $\epsilon_V \ll |\eta_V|$; this is
true also in our model. A spectral index below unity can then only be
realized if the potential is concave, i.e. has a negative second
derivative, leading to a negative (field--dependent) squared mass of the
inflaton. In our model, the latter is given by\footnote{We're neglecting
  the small contribution $\propto \beta$, which is relevant during inflation
  but not afterwards.}
\begin{equation} \label{math:m2-quadratic}
  \tilde{m}^2(\phi) = V"(\phi) = 12 d \left[ (\phi - \frac{2}{3}\phi_0)^2
    - \frac{1}{9} \phi_0^2 \right]\,.
\end{equation}
Thus, the inflaton will become tachyonic if the field lies in the
range
\begin{equation} \label{range}
    \phi_0/3 < \phi < \phi_0\,,
\end{equation}
but has a positive squared mass elsewhere. The minimum value of the
squared mass is
\begin{equation} \label{math:tilde_m_def}
  -\mmin^2 = \min(\tilde{m}^2) = -\frac{4}{3} d \phi_0^2
  = - \frac{1}{3} m_\phi^2\,,
\end{equation}
where
\begin{equation} \label{math:mphi}
  m_\phi = 2 \sqrt{d} \phi_0
\end{equation}
is the inflaton rest mass after the end of inflation (including
now). $m_\phi$ also sets the timescale for inflaton oscillations even
for large amplitude oscillations where the anharmonicity is sizable,
and will therefore be a convenient unit later on.

A tachyonic instability can only occur if the inflaton field is in
the range (\ref{range}). As a first step we show that, in the absence
of backreaction effects, the time the Universe spends in this tachyonic
regime increases with decreasing $\phi_0$. The equation of motion for
the background field, assumed to be spatially constant, is:
\begin{equation} \label{EOM-phi}
  \ddot{\phi} + 3 H \dot{\phi} + V'(\phi) = 0\,.
\end{equation}
Here $H = \dot{a}/a$ is the Hubble parameter. It is given by
\begin{equation} \label{H-bckgd}
H^2 = \frac {\rho_K + \rho_V } {3 \mpl^2}\,,
\end{equation}
where the energy components are
\begin{equation} \label{rho}
\rho_K = \frac{1}{2} \left( \dot{\phi} \right)^2,  \quad \rho_V = V(\phi)\,.
\end{equation}
During inflation, $\rho_K$ is negligible and
$H \simeq H_I = \sqrt{d} \phi_0^2/(3 \, \mpl)$ is essentially
constant; after inflation, $H < H_I$ decreases. The crucial
observation is that $m_\phi / H_I = 6 \mpl/\phi_0$, i.e. the relative
importance of the damping term in eq.(\ref{EOM-phi}) decreases
$\propto 1/\phi_0$.

\begin{figure}[ht]
\centering
\includegraphics[width=0.95\linewidth]{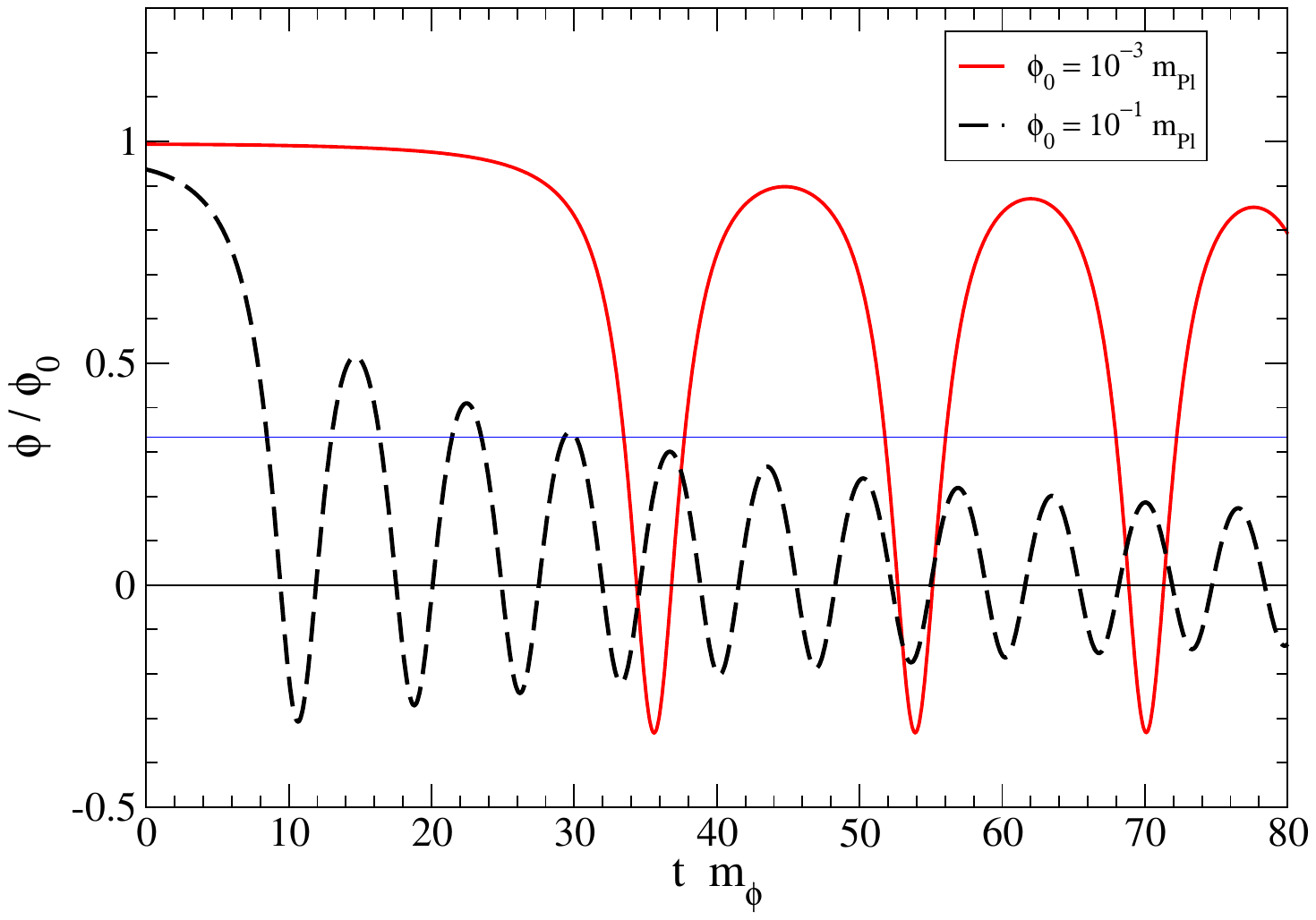}
\caption{Evolution of the (spatially constant) background inflaton
  field according to eqs.(\ref{EOM-phi}) and (\ref{H-bckgd}), for
  $\phi_0 = 0.1 \, \mpl$ (dashed black) and $\phi_0 = 10^{-3} \, \mpl$
  (solid red). The inflaton field is given in units of $\phi_0$, and
  the time in units of $1/m_\phi$, with $t = 0$ denoting the time when
  $\epsilon_V = 0.1$. The inflaton is tachyonic if the field is above
  the thin blue line.}
\label{fig:background}
\end{figure}

This is illustrated in fig.~\ref{fig:background}, which shows the
evolution of the classical inflaton field after the end of slow--roll
inflation; the latter is (somewhat arbitrarily) defined by\footnote{At
  this point $|\eta_V|$ is typically already well above $1$. However,
  since the equation of motion (\ref{EOM-phi}) depends on the first
  derivative of the potential, and thus on $\sqrt{\epsilon_V}$ rather
  than on $\eta_V$, $\phi$ stays close to $\phi_0$ until $\epsilon_V$
  becomes sizable.}  $\epsilon_V = 0.1$, which implies
$\phi = \phi_0 \left( 1 - 0.2 \sqrt{\phi_0/\mpl} \right)$. For
$\phi_0 = 0.1 \,\mpl$ (black dashed curve) the field quickly starts to
oscillate. Because damping is sizable, it re--enters the tachyonic
regime only three times; by $t \, m_\phi = 80$ the oscillations have
become quite harmonic. In contrast, for $\phi_0 = 10^{-3} \, \mpl$
(solid red curve), field oscillations only start at
$t \simeq 30/m_\phi$. Since damping is much less important, the
oscillations remain very anharmonic even at $t = 80/m_\phi$, with the
inflaton field spending most of its time in the tachyonic range
(\ref{range}). In fact, our numerical analysis shows that, if the
dynamics was entirely determined by eqs.(\ref{EOM-phi}) and
(\ref{H-bckgd}), the number of times the inflaton field re--enters the
tachyonic regime would simply scale $\propto 1/\phi_0$.

The evolution of the background field is of interest mainly because it
may allow some perturbations to grow exponentially. This tachyonic
instability comes about if for some wave vectors $\vec{k}$ the
inflaton perturbation $\delta \phi_{\vec{k}}$ develops an imaginary
frequency $\omega_k$. To linear order, $\delta \phi_{\vec{k}}$
satisfies the equation of motion
\begin{equation} \label{EOM-delta-phi}
  \delta \ddot{\phi}_{\vec{k}} + 3 H \delta \dot{\phi}_{\vec{k}}
  + \omega_k^2 (t) \delta \phi_{\vec{k}} = 0\,, \quad {\rm with} \quad
  \omega_k^2 = k^2 / a^2 + \tilde{m}^2(\phi)\,.
\end{equation}
We see that $\omega_k^2 < 0$ when the background field $\phi$ lies in
the range (\ref{range}) and
$k^2 \equiv \vec{k}^2 < - a^2 \tilde{m}^2(\phi)$.

During inflation, $\tilde{m}^2$ is negative but small in magnitude:
$|\eta_V| \ll 1$ implies $\vert\tilde{m}^2\vert \ll H^2$. The tachyonic
instability therefore has negligible effect during the slow--roll
epoch.

In contrast, during the oscillatory phase, i.e. in the reheating
epoch, $m_\phi^2 > H^2$. Setting $H = 0$ to zeroth--order
approximation, the solution of eq.(\ref{EOM-delta-phi}) can
approximately be written as \cite{Sakurai:2011zz, mathews-walker}
\begin{equation} \label{delta-phi-sol}
  \delta \phi_{\vec{k}} = \frac {A_{\vec{k}}} {\sqrt{2\omega_k(t)}}
  \exp[-i\int^t \omega_k(t') \dd{t'}]  + \frac {B_{\vec{k}}}
  {\sqrt{2\omega_k(t)}} \exp[+i\int^t \omega_k(t') \dd{t'}]\,,
\end{equation}
where the coefficients $A_{\vec{k}}$ and $B_{\vec{k}}$ are determined
by the initial conditions. This approximation is only valid when the
frequency $\omega_k(t)$ is changing slowly (or adiabatically),
$|\dot{\omega}| / \omega^2 \ll 1$. For $\omega_k^2 < 0$, the solution
(\ref{delta-phi-sol}) evidently has exponentially decaying and
exponentially growing solutions; clearly the latter will dominate at
late times.

As already noted, the approximate solution (\ref{delta-phi-sol}) has
been derived by setting $H=0$ in the equation of motion. This is
reasonable only for modes well inside the horizon, i.e. for physical
momentum $q = k/a > H$. We saw above that $m_\phi \gg H$, and hence
$|\tilde m_{\rm min}| \gg H$, if $\phi_0 \ll \mpl$; exponential growth
can then occur for $|\tilde m_{\rm min}| > q > H$. Note also that for
an effectively matter--dominated universe, $H \propto
a^{-3/2}$ decreases faster than the physical momentum $q \propto a^{-1}$,
i.e. a mode that is inside the Hubble horizon at the onset of inflaton
field oscillations will remain inside the horizon.

\begin{figure}[t]
\centering
\includegraphics[width=0.95\linewidth]{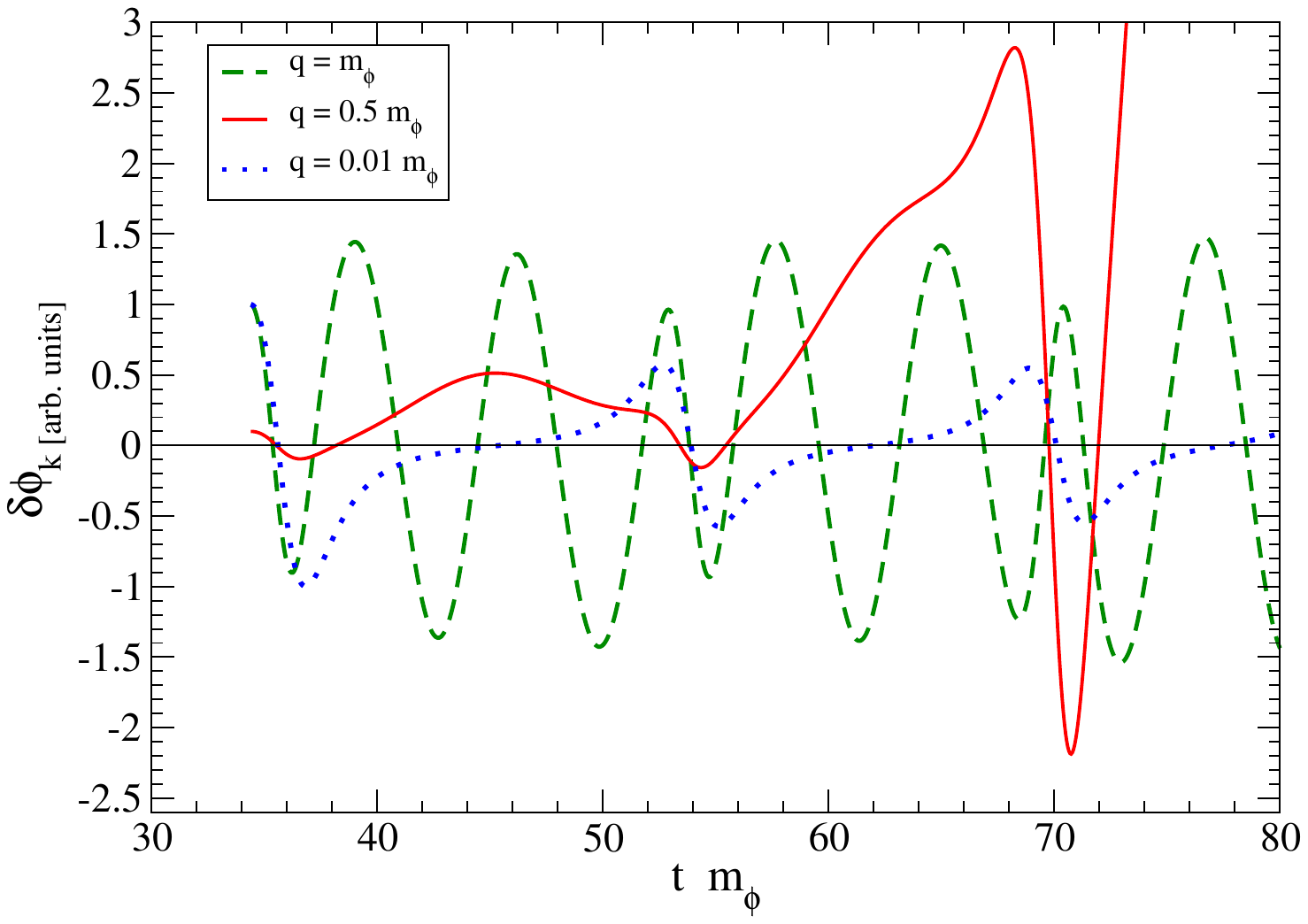}
\caption{Evolution of some Fourier modes of the inflaton field for
  $\phi_0 = 10^{-3} \, \mpl$. $q$ is the absolute value of the
  physical wave vector at the time when the background field first
  crosses zero. At this ``initial'' time, the time derivative of the
  perturbation has been set to zero; $\delta \phi_k$ has been set to
  $1$ for $q = m_\phi$ (dashed green) and for $q = 0.01 \, m_\phi$
  (dotted blue), but to $0.1$ for $q = 0.5 \, m_\phi$ (solid
  red). Since the evolution equation (\ref{EOM-delta-phi}) is linear
  in $\delta \phi_k$, its initial value simply scales the curves up or
  down.}
\label{fig:modes}
\end{figure}

Fig.~\ref{fig:modes} shows some numerical solutions of
eq.(\ref{EOM-delta-phi}) for $\phi_0 = 10^{-3} \, \mpl$; the evolution
of the background field has again be computed from eqs.(\ref{EOM-phi})
and (\ref{H-bckgd}). We have initialized the Fourier modes at the
onset of oscillations, defined by the time when the background field
first crosses zero; $q$ is the physical momentum at that time. For
simplicity we have set the initial time derivative $\delta \dot\phi_k$
to zero. Since eq.(\ref{EOM-delta-phi}) is linear, the solution is
then proportional to the initial value. We see that the mode with
$q = m_\phi$ (dashed green curve), which is outside the tachyonic
window, oscillates with an amplitude that stays roughly constant over
the range of time shown; eventually it gets damped by the Hubble
expansion. The mode with $q = 0.01 \, m_\phi$ (blue dotted curve) does
have a tachyonic mass; however, since its initial wavelength is
comparable to the Hubble radius, it also does not show significant
growth. In contrast, the node with $q = 0.5 m_\phi$ (solid red) grows
very quickly, each maximum being about a factor of $5$ above the
previous one. This confirms our qualitative discussion above.

\section{Floquet analysis}
\label{sec:floquet}

In this section we investigate the exponential growth of some modes
more quantitatively. We stay in the linear regime where the field
perturbation $\delta \phi$ is small compared to the background field
$\expval{\phi}$. This Floquet analysis will also be helpful for
setting the lattice parameters later on. Moreover, it will allow us to
estimate the time when the back--reaction of the perturbation modes on
the evolution of the background field becomes important; beyond this
point the perturbative analysis will no longer work. Finally, we'll
present a simple analytical model which confirms our numerical
results.

\subsection{Linearized equation for field fluctuations}

The goal is to solve equation \eqref{EOM-delta-phi} for given
$k$. More exactly, we want to identify the exponentially growing
modes, and estimate just how fast they grow. To this end we perform
a Floquet analysis. The Floquet theorem states that for equations of
motion with periodic frequency $\omega_k(t)$, the solutions have the
form
\begin{equation} \label{floquet-solution-general}
	y_k (t) = A(t) e^{\mu_k t}  + B(t) e^{-\mu_k t},
\end{equation}
with $A(t)$ and $B(t)$ being periodic functions and $\mu_k$ a
momentum--dependent complex number: the Floquet exponent. Evidently
modes with $\realp(\mu_k) \neq 0$ will experience exponential growth; these
modes might therefore ultimately affect the background field evolution
\cite{aminNonperturbativeDynamicsReheating2015}.

For the Floquet theorem to work, there should be no damping term and
the frequency $\omega^2_k(t) = k^2/a^2 + \tilde{m}^2 (\phi)$ needs to
be strictly periodic (in time). Thus, we have to set
$H(t) \rightarrow 0$ everywhere, including in the evolution of the
background field, which then becomes perfectly periodic. In our case
this can be a good approximation only for $\phi_0 \ll \mpl$, where the
amplitude of the background field oscillation indeed decreases very
slowly as we saw in fig.~\ref{fig:background}. The effective mass
$\tilde{m}^2$ is thus computed from the numerical solution of
eq.(\ref{EOM-phi}) with $H = 0$. We chose the initial conditions such
that the oscillation amplitude matches the amplitude of the first
``period'' of the exact solution of eq.(\ref{EOM-phi}), including the
damping term. This trick allows us to match more closely the real
post--inflationary dynamics of our mode.

The basic idea of the Floquet analysis is to first convert the second
order differential equation \eqref{EOM-delta-phi} into two first order
differential equations, which can be written in matrix form as
\begin{equation}
	\partial_t x(t) = U(t) x(t)\,.
\end{equation}
The solutions are encoded in a fundamental matrix
$\mathcal{O}(t, t_0)$, with $O(t_0,t_0)$ being the identity matrix and
$x(t) = O(t,t_0) x(t_0)$. By the Floquet theorem, the eigenvalues of
the monodromy matrix $\mathcal{O}(t_0 + T, t_0)$ are the Floquet
exponents. Here $T$ is the period of the system; in our case
$T \sim 1/m_\phi$, and the Floquet exponents depend on $\phi_0$ and
$q$. Details of computation are given in appendix
\ref{sec:app-floquet}.

\subsection{Numerical result}

\begin{figure}[ht]
\centering
\includegraphics[width=0.75\linewidth]{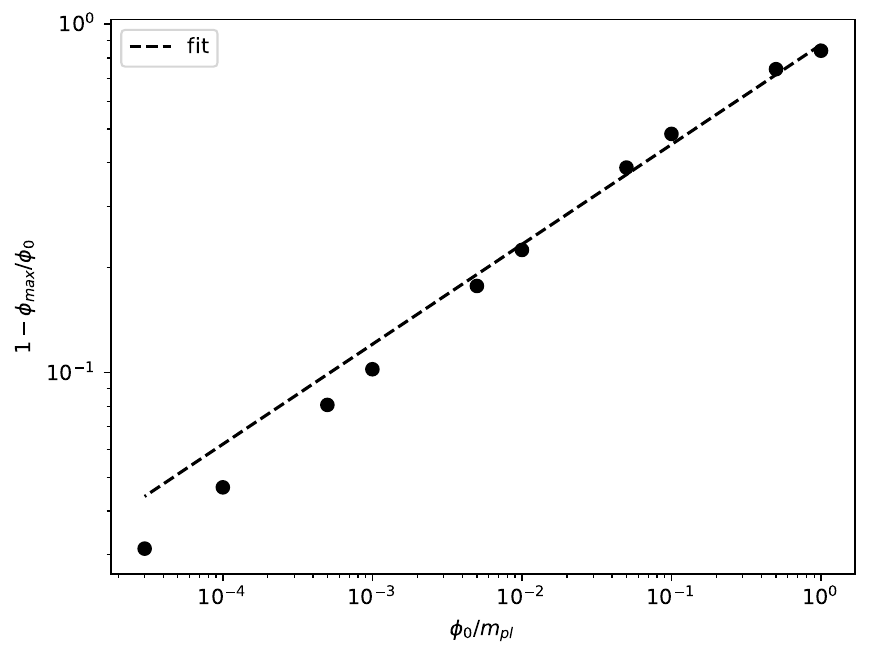}
\caption{The ``loss'' of amplitude during the first oscillation in the
  exact equation of motion for the background field $\phi$. The fit
  function is
  $1-\phi_{\text{max}}/\phi_0 = 0.87 \cdot (\phi/m_{\pl})^{0.29}$.}
\label{fig:1st_peaks}
\end{figure}

As mentioned in the previous section, the background field
$\expval{\phi}$ is numerically calculated without the damping
term. The initial amplitude is set to equal the amplitude of the first
oscillation of the numerical solution of eq.(\ref{EOM-phi}) including
the damping term, i.e.  the value $\phi_{\text{max}}$ of the first
maximum after the inflaton field first crossed
zero. Fig.~\ref{fig:1st_peaks} shows numerical results for this
quantity, as well as a convenient fit function which we use in our
numerical results presented below.

\begin{figure}[ht]
\begin{subfigure}[t]{0.5\textwidth}
\begin{center}
\includegraphics[width=\linewidth]{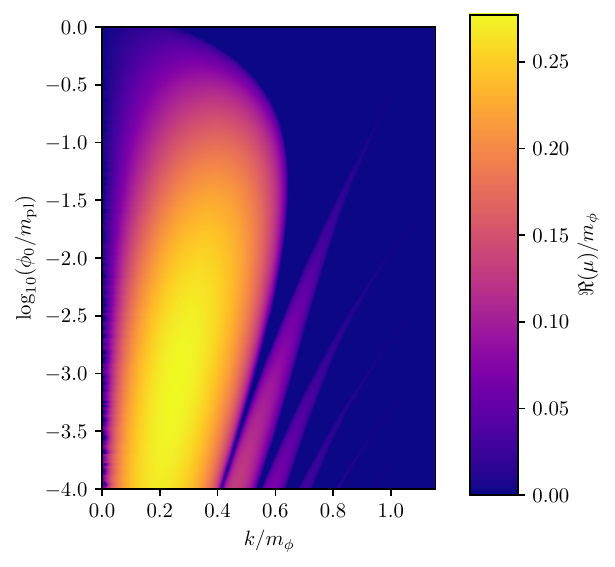}
\end{center}
\caption{Floquet coefficients. The frequency $\omega(t)^2$ in equation
  \eqref{EOM-delta-phi} can be negative for modes with
  $k < m_\phi / \sqrt{3} \approx 0.58 m_\phi$; note that $H=0$ implies
  that $k=q$, i.e. there is no difference between co--moving and
  physical momenta in this approximation.}
\label{fig:floquet-map}
\end{subfigure}
\begin{subfigure}[t]{0.5\textwidth}
\begin{center}
\includegraphics[width=\linewidth]{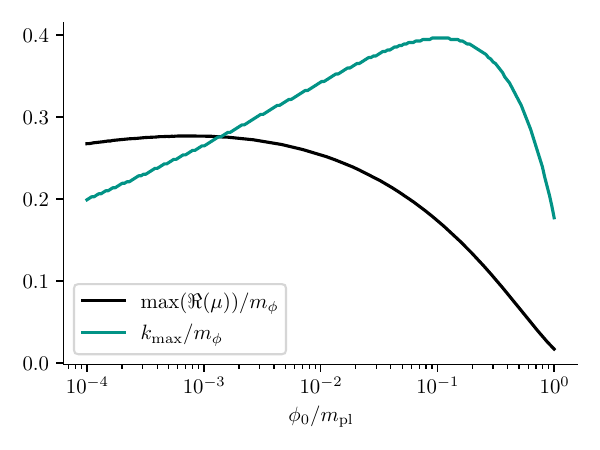}
\end{center}
\caption{Location and height of $\max(\realp(\mu))$.}%
\label{fig:floquet-max}
\end{subfigure}
\caption{Numerical results of Floquet analysis.}
\label{fig:floquet}
\end{figure}

Results for the Floquet exponent with the largest real part are shown
in fig.~\ref{fig:floquet}. We see in fig.~\ref{fig:floquet-max} that
for $\phi_0 \geq 10^{-3} \, \mpl$, smaller $\phi_0$ leads to larger
maximal $\realp(\mu)$, thus faster growth. This saturates for even
smaller $\phi_0$, since then the loss of amplitude in the first
oscillation is basically negligible, as shown in
fig.~\ref{fig:1st_peaks}. In fact, the Floquet analysis isn't really
applicable to $\phi_0 \gtrsim 0.1 \, \mpl$, since then the background
field re--enters the tachyonic window only a few times during
oscillation, with quickly decreasing amplitude, as we saw in
fig.~\ref{fig:background}; keeping this amplitude fixed, as we do
here, is then a bad approximation.

We also see that the Floquet coefficient reaches its maximum for
smaller $k$ when $\phi_0$ is reduced. This allows the appearance of
additional instability bands at larger $k$ for
$\phi_0 < 10^{-2} \, \mpl$, as shown in fig.~\ref{fig:floquet-map}.

Another important feature of the Floquet map is that
$\mu_{k=0} \rightarrow 0$. Physically, this can be understood since
the background field itself, which corresponds to the homogeneous part
with $k=0$, should not get affected by the instability.
Mathematically $k \neq 0$ is required in order to create a mismatch
between the period of the perturbation and that of the background
field. A nice mathematical explanation has been given in
\cite{Karam:2021sno}.

\subsection{Time for back-reaction}\label{sec:br-time}

The energy source for the growth of perturbations is the spatially
homogeneous, oscillating background field, see
eq.\eqref{EOM-delta-phi}. So far we have assumed that the
perturbations do not change the evolution of the background
field. This is not longer true as soon as some perturbations
$\delta \phi_k$ become large. Physically it is clear that the
exponential growth of the perturbations cannot continue once most of
the energy initially stored in the background field has been
transferred to the $\delta \phi_k$.

Here we wish to estimate the time when the back--reaction becomes
important. To that end we again ignore Hubble friction. We estimate
the timescale of back--reaction from the maximal Floquet exponent
$\realp(\mu_k)$, assuming it to be constant throughout and ignoring
that the physical momenta can pass through the instability
band(s). This will tend to over--estimate the growth of perturbations
compared to a full numerical analysis, and thus to under--estimate the
time when back--reactions become important.

This time scale depends not only on the growth rate of perturbations,
but also on their size at the end of inflation (more precisely, at the
time when the background field begins to oscillate around the
minimum). As shown in \cite{Polarski:1995jg, Khlebnikov:1996mc} one
can use the standard Bunch--Davis vacuum to fix the initial condition
for the fluctuations, at least for quickly growing modes where the
occupation number will become large, a posteriori justifying a
(semi--)classical treatment. The time scale $\Delta t_{\rm br}$ for
back--reactions can thus be estimated from 
\begin{equation} \label{math:back-react-cond}
\frac{k}{2\pi} \exp(\realp(\mu_k) \Delta t_{\rm br}) \stackrel{!}{\simeq} \phi_0
\,,
\end{equation}
for $k = k_{\rm max}$ where $\realp(\mu_k)$ has its maximum. This implies
\begin{equation} \label{math:back-react-time}
  \Delta t_{\rm br} \simeq \frac{1}{\realp(\mu_k)} \ln \left( \frac {2\pi \phi_0}
    {m_\phi} \frac {m_\phi} {k_{\rm max}} \right) = \frac{1}{\realp(\mu_k)}
  \left[ 18.6 + \ln(10) 
    \log_{10}  \left(\frac{\mpl}{\phi_0} \right) -
    \ln(\frac{k_{\text{max}}}{m_\phi}) \right]\, .
\end{equation}
In the second step we have used eqs.(\ref{math:mphi}) and
(\ref{par-values}). Using results from the right frame of
fig.~\ref{fig:floquet}, we see that for $\phi_0 \lesssim 0.1 \mpl$ it
takes (at least) roughly $100 \ m_{\phi}^{-1}$ for back--reactions to
become important. However, recall from fig.~\ref{fig:background} that
for $\phi_0 = 0.1 \mpl$ the background field no longer reaches the
instability band at $t = 100\ m_\phi^{-1}$, once Hubble friction is
included. We thus conclude that non--linear effects in the evolution
of the inflaton field will become important only for
$\phi_0 < 0.1 \mpl$.

Recall that our Floquet analysis ignores the Hubble expansion. The
reduction of the amplitude of the oscillations of the background field
should be essentially irrelevant if $\realp(\mu_k) \gg H$. Since
\begin{equation} \label{math:mu-over-H}
  \frac{\realp(\mu_k)}{H} \gtrsim  \frac{\realp(\mu_k) / m_\phi} {H_I/m_\phi} = 6
  \frac{\realp(\mu_k)}{m_\phi} \left( \frac{\phi_0}{\mpl} \right)^{-1},
\end{equation}
this condition is satisfied for momenta near the maximum of the instability
band, where $\realp(\mu_k) \geq 0.2 m_\phi$, as long as $\phi_0 < 0.1 \mpl$.
Similarly, the redshift changes physical momenta significantly only after
about a Hubble time. The approximation of constant $q$ should thus
hold over the relevant period of time if $H < m_\phi/100$, which is true
for $\phi_0 < 0.05\ \mpl$. In this range of $\phi_0$ we therefore expect
non--linear dynamics to become important quickly, in practice after
${\cal O}(10)$ oscillations of the background field.

\subsection{Analytical treatment}

Before turning to a fully numerical treatment of the nonlinear
dynamics, we want to better understand some features of the Floquet
map.  In particular, we want to qualitatively discuss the dependence
of the Floquet exponent on the model parameter $\phi_0$. The usual way
to analytically compute the Floquet exponential is by treating the
non--linearities as perturbations
\cite{aminInflatonFragmentationEmergence2010}. This does not work in
our model since the potential is specifically engineered so that all
terms are equally important for inflaton field values leading to a
negative effective inflaton mass.

Instead we fit $\tilde{m}^2 (\phi)$ in eq.\eqref{EOM-delta-phi}
directly as a sum of delta functions and (possibly) step functions
\cite{Koivunen:2022mem}. Then one can use continuity conditions in
order to calculate the Floquet exponent. Our ansatz reads:
\begin{equation} \label{math:Koivunen-omega2}
  \omega_k^2 = -\Gamma_k^2 + \sum_{j \in \Z} f(t-jT) =
  k^2 - \Gamma_0^2 + \sum_{j \in \Z} f(t-jT)\, ,
\end{equation}
where $T$ is the period of the oscillation of the background
field. For the ``delta model'', the function $f(t)$ is simply
\begin{equation} \label{math:fdelta}
	f(t) = \Lambda \delta(t)\,,
\end{equation}
while for the ``box model'' we assume
\begin{equation}
  f(t) = (\tilde{\omega}_0^2 - \Gamma_0^2) \theta(T_1/2 - |t|)
  - \tilde{\Lambda} \delta(|t| - T_1/2)\,.
\end{equation}
The coefficients $\Lambda$ and $\tilde{\Lambda}$ will be fixed through
the consistency condition $\mu_{k=0} \rightarrow 0$, which is required
in order to obtain the correct shape of the Floquet map. $\Gamma_0^2$
is a constant negative squared mass term; it is determined by taking
the average of leftmost and rightmost points of one single period of
$\tilde{m}^2 (t)$. In the box model, the length ($T_1$) and height
($\tilde{\omega}_0^2$) of the box are determined by a least square
fit.\footnote{Since a perfect step function has infinite derivative,
  many curve fitting routines will have difficulties. To circumvent
  this, we use the sigmoid function to mimic the logistic function
  with steepness parameter $\kappa = 500$.} In turn, during the (possibly) tachyonic
phase $\omega_k^2 = -\Gamma_k^2$ as in the delta model; evidently
$\omega_k^2 < 0$ for $k^2 < \Gamma_0^2$. During the non--tachyonic
phase, $\omega_k^2 = k^2 + \tilde{\omega}_0^2 > 0$.

\begin{figure}[ht]
\centering
\includegraphics[width=0.8\textwidth]{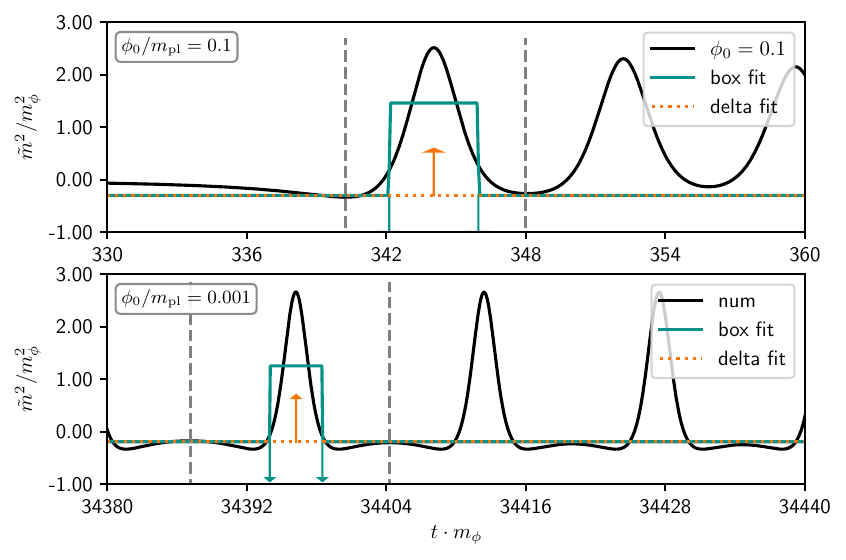}
\caption{The smooth black lines show the effective squared inflaton
  mass $\tilde{m}^2 (t)$ with $\phi_0 = 0.1 \ \mpl$ (top) and
  $\phi_0 = 0.001 \ \mpl$ (bottom), computed from solutions of the
  ordinary differential equation \eqref{EOM-phi} with $H=0$. In the
  delta model this is represented by the sum of a negative constant
  plus a delta--function at the location of the maximum of the
  effective mass (orange arrows). In the box model (blue) the fit
  jumps from a finite negative to a finite positive value, and
  subtracts delta--functions at the transition points in order to
  reproduce the dips beside the central peak(blue arrows). The fits
  are performed in the region enclosed by the gray dashed lines.}
\label{fig:flo-m2}
\end{figure}

Numerical results for $\tilde m^2$ and its representation in the two
models are shown in fig.~\ref{fig:flo-m2} for two values of
$\phi_0$. We see that for $\phi_0=0.001 \ \mpl$ the positive peaks are
much narrower than for $\phi_0 = 0.1 \ \mpl$; this agree with
fig.~\ref{fig:background}, where the inflaton field spent much more
time in the tachyonic region for smaller $\phi_0$. This leads us to
expect that the delta fit should work better for
$\phi_0 = 0.001 \ \mpl$ than for $\phi_0 = 0.1\ \mpl$. In turn, the
box fit might work better for $\phi_0 / \mpl = 0.1$.

The Floquet exponent of the delta model is given by
\begin{align} 	\label{math:mu-delta}
  \mu_k = \frac{1}{T} \realp \; \arcosh \left[ \cosh(\Gamma_k T) -
  \frac{\Lambda}{2\Gamma_k} \sinh(\Gamma_k T) \right], \quad
  \Lambda = 2 \Gamma_0 \coth(\frac{\Gamma_0 T}{2})\,;
\end{align}
for the box model we find
\begin{subequations} \label{math:mu-box}
\begin{align}
\begin{split}
  \mu_k &= \frac{1}{T} \realp\; \arcosh \Bigg\{ [\cos(T_1 \tilde{\omega}_k )
  + \frac{\tilde{\Lambda}}{\tilde{\omega}_k} \sin(T_1 \tilde{\omega}_k)]
  \cosh(\Gamma_k T_2) \\
  & \quad + \left[ \frac{\Gamma_k^2 - \tilde{\omega}^2 + \tilde{\Lambda}^2}
    {2\Gamma_k \tilde{\omega}_k} \sin(T_1 \tilde{\omega}_k)
    + \frac{\tilde{\Lambda}}{\Gamma_k} \cos(T_1 \tilde{\omega}_k) \right]
  \sinh(\Gamma_k T_2) \Bigg\}\,, \\
\end{split} \\
  \tilde{\Lambda} &= \tilde{\omega}_0 \tan(\frac{T_1 \tilde{\omega}_0}{2})
                    - \Gamma_0 \coth(\frac{T_2 \Gamma_0}{2})\,,
\end{align}
\end{subequations}
with $T_2 = T - T_1$.

\begin{figure}[ht]
\centering
\includegraphics[width=0.8\textwidth]{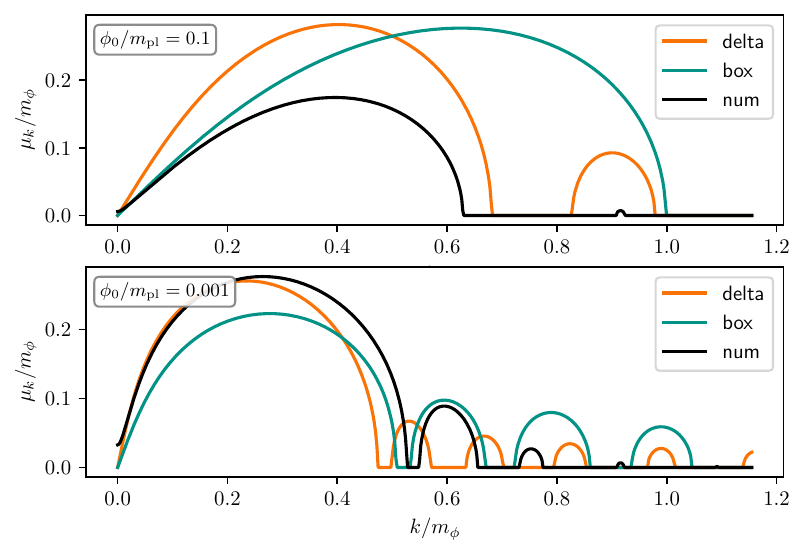}
\caption{Floquet exponents as a function of $k$ for
  $\phi_0 = 0.1 \ \mpl$ (top) and $\phi_0 = 0.001 \ \mpl$
  (bottom). The black curves show numerical results, basically slices
  through the left frame of fig.~\ref{fig:floquet}. The orange and
  blue curves show analytical results based on the delta model,
  eq.(\ref{math:mu-delta}), and the box model, eq.(\ref{math:mu-box}),
  respectively.}
\label{fig:flo-ana}
\end{figure}

Examples of these analytical results are compared in
fig.~\ref{fig:flo-ana}.  to numerical results. For
$\phi_0 = 0.1 \ \mpl$ (top frame) both analytical approaches
overestimate $\realp(\mu_k)$, but the delta model gets the location of
the maximum $k_{\rm max}$ roughly right. The bad performance of the
box model in this case might be due to the fact that the exact
$\tilde m^2$ actually does not show pronounced dips close to the
central peak in this case. On the other hand, both models describe the
main instability band fairly well for $\phi_0 = 0.001 \ \mpl$. The
delta model does a little better here, but the box model also
reproduces the second instability band quite accurately; this has also
been observed in \cite{Koivunen:2022mem}. Both models reproduce the
trend seen in fig.~\ref{fig:floquet} that reducing $\phi_0$ reduces
$k_{\rm max}$ and increases the number of instability bands.

\section{Lattice simulation}
\label{sec:lattice}

In order to accurately describe the non--linear dynamics we utilize
lattice simulations. In the following subsection we describe the setup
of our simulation. Focusing on $\phi_0 = 0.01 \ \mpl$, in subsequent
subsections we analyze the behavior of average field quantities, the
Hartree approximation for the average field, the spectrum of
fluctuations, and oscillons. In the final subsection we discuss the
dependence on $\phi_0$.

\subsection{Simulation setup}
\label{sec:lattice-setup}

We mainly use the program \texttt{CosmoLattice} \cite{figueroaCosmoLattice2021}
to simulate the dynamical evolution of the classical inflaton field. Some
modifications to the program were made, but the core of the program
stays intact. The program employs the following dimensionless variables:
\begin{equation} \label{math:vardef}
  \tilde{\phi} := \frac{\phi}{f_*}, \quad
  \dd{\tilde{\eta}} := a^{-\alpha}(\tilde{\eta}) \omega_* \dd{t}, \quad
  \dd{\tilde{x}^i} := \omega_* \dd{x^i}\,.
\end{equation}
Here $t$ is the cosmic time and the $x^i$ are comoving coordinates.
In general the values of $f_*$, $\omega_*$ and $\alpha$ affect the
numerical stability. We set $f_*$ to $\phi_0$, which sets the order of
magnitude of the oscillating background field at the beginning of the
simulation. $\omega_*$, which sets the unit of time and spatial
distance and thus momentum, is chosen so that the oscillation period
of inflaton is roughly of order of unity. The dynamics of inflation is
often analyzed using conformal time, which corresponds to
$\alpha = 1$. However, we are interested in the post--inflationary
epoch. A simple oscillating field can easily be analyzed using the usual cosmic
time, $\alpha = 0$.  In the interesting field range all three terms in
the inflaton potential contribute about equally, i.e. the oscillation
is highly anharmonic, as we saw in fig.~\ref{fig:background}.
Nevertheless we find that setting $\alpha = 0$ does not lead to any
numerical late--time instability. Hence, we set
\begin{equation} \label{math:lattice-var}
  f_* =  \phi_0\,, \quad
  \omega_* = \frac{\sqrt{d}}{3} \phi_0\,, \quad
  \alpha = 0\,.
\end{equation}
We checked that setting $\alpha=1$ leads to identical result.  Our
choice of $f_*$ and $\omega_*$ implies that the dimensionless
potential energy
\begin{equation} \label{math:Vtilde}
\tilde{V} = \frac{V(\phi)}{f_*^2 \omega_*^2}
\end{equation}
is basically independent of $\phi_0$. In order to avoid confusion, we
will stick with $m_\phi = 6 \omega_*$ as unit for length, time, and
momentum when presenting results of the lattice simulation, just as we
did in the previous chapters.

As discussed in Sec.~3.3, the initial fluctuations in momentum space
are set according to the usual Bunch--Davis vacuum. Since we are
working on a lattice with finite distance $\delta x$ between
neighboring grid points we must set a UV cutoff (\verb|kCutoff|) for
the initial fluctuations, beyond which they are set to zero.  This
``external'' cutoff should be smaller than the implicit lattice cutoff
$2\pi/\delta x$. Another important parameter is the overall size of
the lattice, or the number of grid points (in one direction) $N$. For
convenience, we set the infrared cutoff \verb|kIR| instead of $\delta x$
in the program input; they are related by
\begin{equation} \label{math:lattice-spacing}
\delta x = \frac{2 \pi }{N} \frac{1}{\texttt{kIR}}\,.
\end{equation}
The UV side of power spectra must always be checked at the end of the
simulation. If \verb|kCutoff| is too large, due to the discrete nature
of lattice and thus periodicity in the reciprocal lattice, any
dynamically generated super--UV modes get ``reflected'' to the
infrared; this can cause a spurious enhancement at large wavelengths
(small $k$).

We always use the velocity verlet method of order 4 (VV4)
implemented in \cite{figueroaCosmoLattice2021} to solve
the differential equations. The accuracy of the simulation is
quantified, as implemented in \cite{figueroaCosmoLattice2021}, by the
relative difference of both sides of the second Friedmann equation,
derived from the first Friedmann equation
\begin{equation} \label{math:friedman}
  H^2 = \frac {\rho_K + \rho_G + \rho_V} {3 \mpl^2},
\end{equation}
and from the $ij$-component of the Einstein equations:
\begin{equation} \label{math:lat-cons}
  \frac{a''}{a} = \frac{a^{2\alpha}}{3} \left( \frac{f_*}{\mpl} \right)^2
  \left[ (\alpha - 2) \expval{\tilde{\rho}_K}  + \alpha \expval{\tilde{\rho}_G}
    + (\alpha + 1) \expval{\tilde{\rho}_V} \right]\,.
\end{equation}
$\rho_K$ and $\rho_V$ in eq.\eqref{math:friedman} have already been
defined in eq.\eqref{rho}, and $\rho_G = (\nabla \phi)^2 / (2 a^2)$. In
eq.(\ref{math:lat-cons}), $\expval{...}$ denotes a volume
averaged quantity, a prime denotes a derivative w.r.t. $\tilde \eta$,
and the $\tilde{\rho}_i$ are the dimensionless energy densities
defined via the rescaled variables of eq.\eqref{math:lattice-var}:
\begin{equation} \label{math:rho-tilde}
  \tilde \rho_K = \frac{1}{2} \left( \tilde\phi' \right)^2 \,, \quad
  \tilde \rho_G = \frac{1}{2} \left( \tilde\nabla \tilde\phi \right)^2\,,
  \quad \tilde \rho_V = \tilde V\,.
\end{equation}

For most of this chapter we will set
\begin{equation} \label{math:phi0}
  \phi_0 = 0.01 \ \mpl\,;
\end{equation}
the preceding discussion shows that this is one of the largest values
of $\phi_0$ where we can expect full--fledged nonlinear dynamics. We
first tried starting our simulation at the (canonical) end of
slow--roll inflation, where the second slow--roll parameter
$|\eta_V| = 1$.  The initial time derivative of the background field,
$\dot{\phi}$, is computed from its equation of motion
(\ref{EOM-phi}). However, we noticed that $\epsilon_V \ll 1$ for
several more e--folds, which means that the fluctuations still play no
role and just redshift away. We therefore switched to starting our
lattice simulation at the time when $\ddot a = 0$, computed from the
numerical solution of the equation of motion (\ref{EOM-phi}). By the
second Friedmann equation \eqref{math:lat-cons}, this is equivalent to
$(\dot{\phi})^2 = V(\phi)$; this happens slightly before the
background field first crosses zero. Our final choices of the lattice
parameters for different values of $\phi_0$ can be found in
appendix~\ref{sec:app-lat-para}.

Using eq.(\ref{math:lat-cons}) as described above, we checked that the
numerical error is kept within $10^{-8}$ throughout the
simulation. However, since the expansion of the universe reduces the
implicit UV cutoff on physical momenta we cannot completely avoid the
problem of ``reflecting'' modes with large $k$ into the IR
region. Moreover, \texttt{CosmoLattice} does not include the
backreaction of field inhomogeneities onto the metric, which is
assumed to be given by the Friedman--Robertson--Walker form
throughout. This means that the growth of overdensities due to
gravitational attraction is not included. Due to these issues, the
simulation eventually becomes physically unreliable, even if numerical
errors remain small. We will come back to this later.

Note that we do not include the decay of the inflaton to other
particles in our simulation. This is well justified if the inflaton
couplings to other particles are small enough that they do not distort
the inflaton potential through radiative corrections. As shown in
\cite{dreesSmallFieldPolynomial2021} this implies
$\Gamma_\phi < 10^{-14} m_\phi$ for $\phi_0 = 0.01 \; \mpl$, where
$\Gamma_\phi$ is the total (perturbative) decay with of the inflaton;
for smaller $\phi_0$ the upper bound on $\Gamma_\phi / m_\phi$ is even
smaller. Hence inflaton decay is certainly negligible during our
simulation, which covers $t \lesssim 10^3/m_\phi$.

We are now ready to present some numerical results. We begin with a
discussion of volume--averaged quantities.

\subsection{Average quantities}
\label{subsec:average}

\begin{figure}[ht]
\centering
\includegraphics[width=0.75\linewidth]
{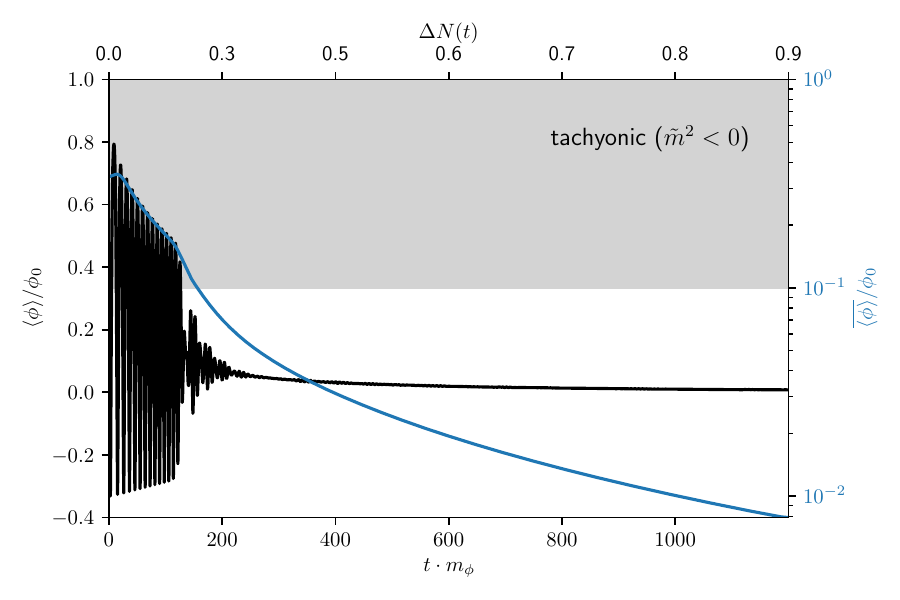}
\caption{The black curve shows the evolution of the volume--averaged
  value of the inflaton field and refers to the linear left
  $y-$axis. The blue curve refers to the logarithmic right $y-$axis
  and gives the volume--averaged field value averaged over roughly two
  periods using a Gaussian filter. Two horizontal axes are given: one
  is the usual cosmic time $t$ and the other is the number of e--folds
  after slow--roll inflation, i.e. it shows the evolution of the scale
  factor $a$. The tachyonic region, where the effective mass squared 
  is negative, is shaded grey.}%
\label{fig:0.01-mean}
\end{figure}

We begin with fig.~\ref{fig:0.01-mean}, which shows the
volume--average of the inflaton field as a function of time (black)
and additionally averaged over a couple of oscillations (blue). We see
that at first the field oscillates as predicted by the solution of the
equation of motion (\ref{EOM-phi}). During this epoch the amplitude of
the oscillation, and the time--averaged field, decrease roughly as a
negative power of time. This epoch lasts until
$t \simeq 150 m_{\phi}^{-1}$, which roughly matches the expectation of
sec.~\ref{sec:br-time}. Then the mean field collapses, i.e.  the
oscillation amplitude decreases very quickly. Immediately after this
collapse the volume--averaged field behaves more chaotically at short
times scales, but its time--average continues to decline smoothly.

\begin{figure}[h!]
\centering
\includegraphics[width=0.5\linewidth]
{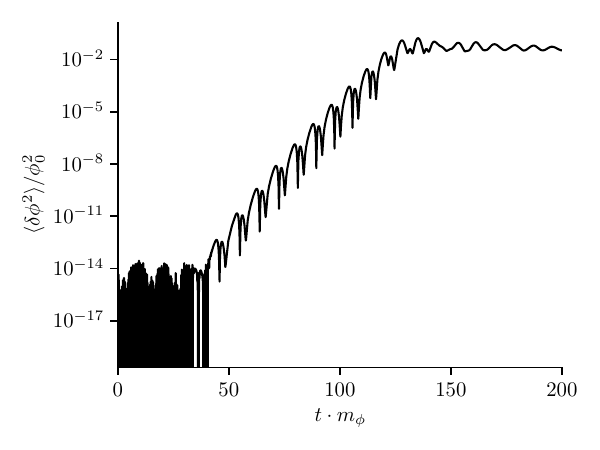}%
\includegraphics[width=0.5\linewidth]
{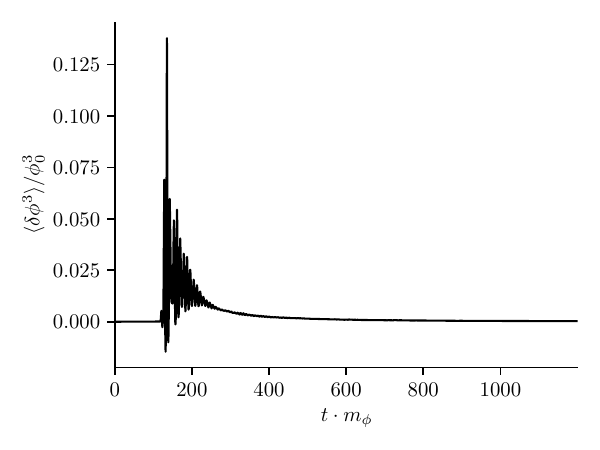}
\caption{Variance and skewness of the inflaton field $\phi$ as function
  of cosmic time.}%
\label{fig:0.01-delta-phi-2-3}
\end{figure}

In order to gain insight into the collapse, we show the time evolution
of the variance and skewness of the inflaton field in
fig.~\ref{fig:0.01-delta-phi-2-3}. The variance, defined as
\begin{equation} \label{math:variance}
\expval{\delta \phi^2} = \expval{\phi^2} - \expval{\phi}^2,
\end{equation}
describes the overall magnitude of the spatial variation of $\phi$. The
left frame of fig.~\ref{fig:0.01-delta-phi-2-3}, which uses a
logarithmic $y-$axis, shows that it is initially very small, although
growing roughly exponentially as expected from our Floquet
analysis. For $50 \leq t \cdot m_\phi \leq 120$ the exponent is about
$0.31/m_\phi$, which is less than twice the maximal Floquet coefficient shown in
fig.~\ref{fig:floquet}; the factor of two arises since the variance is
quadratic in the perturbation $\delta \phi$. 
We note that the variance oscillate along with the exponential growth.
It is surprising, since the initial fluctuations are randomly seeded 
\cite{figueroaCosmoLattice2021}. The same oscillation has been observed 
in \cite{desrochePreheatingNewInflation2005, kainulainen2024tachyonicproductiondarkrelics}
as well. At $t \sim 150 m_\phi^{-1}$, it reaches a maximum at
$\sqrt{\expval{\delta \phi^2}} /\phi_0 \sim 0.4$; we saw in
fig.~\ref{fig:0.01-mean} that just before this time the amplitude of
the oscillating $\expval{\phi}$ was only slightly above, and the
time--averaged field somewhat below, this value. This means that in
some regions of the lattice field fluctuations exceed the background
field. Clearly the fluctuations can no longer be ignored at, and
after, this time.

The skewness $\expval{\delta \phi^3}$ is defined as
\begin{equation} \label{math:skewness}
  \expval{\delta \phi^3} = \expval{ \left( \phi - \expval{\phi} \right)^3 }
  = \expval{\phi^3} - 3 \expval{\phi} \left( \expval{\phi^2} -
    \expval{\phi}^2 \right) - \expval{\phi}^3\,.
\end{equation}
The right frame of fig.~\ref{fig:0.01-delta-phi-2-3} shows that the
skewness is positive most of the time, i.e. the spatial distribution
of the field is skewed towards positive field values. This is
expected, since the choice of a negative cubic coefficient in the
inflaton potential (\ref{math:V-sfpi}) implies
$V(|\phi|) < V(-|\phi|)$.  Note also that if the field fluctuation was
purely Gaussian, any odd moment of the field, and hence also the
skewness, would vanish. The non--zero three--point correlation here
originates from the non--linear interaction.\footnote{Non--Gaussianity
  can also be tested by the kurtosis
  $\expval{\phi^4}/\expval{\phi^2}^2$. For purely Gaussian
  fluctuations this always equals $1$
  \cite{felderNonlinearInflatonFragmentation2007}.} By using a linear
$y-$axis in this figure, we demonstrate just how rapid the onset of the
non--linear regime is.

\begin{figure}[ht]
\centering
\includegraphics[width=0.5\textwidth]
{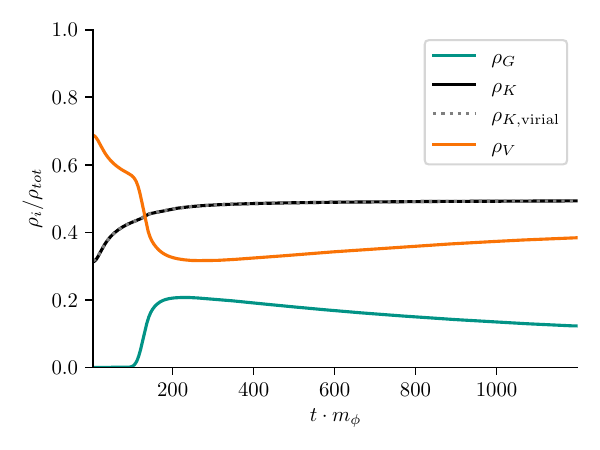}%
\includegraphics[width=0.5\textwidth]
{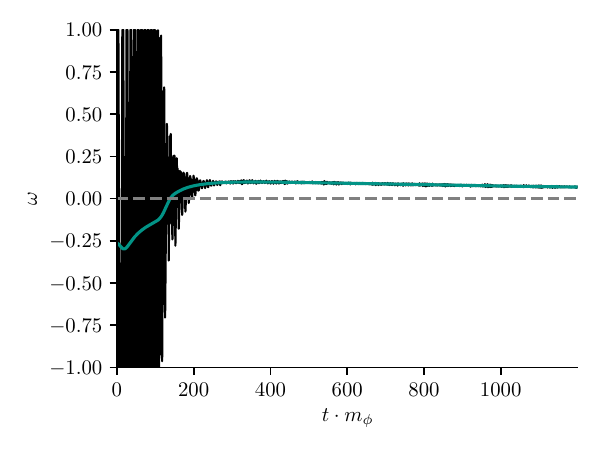}
\caption{The left frame shows the potential energy $\rho_V$ (red),
  kinetic energy $\rho_K$ (black) and gradient energy $\rho_G$ (blue)
  in units of the total energy, smoothed over roughly two periods of
  oscillations of the background field; the dotted gray curve (almost
  coinciding with the black curve) shows the virial prediction for the
  kinetic energy, eq.(\ref{math:virial}). The right frame shows the
  equation of state parameter $\omega$ (black) and its value smoothed
  over two periods (blue).}
\label{fig:0.01-w-energy-frac}
\end{figure}

This can also be seen from the left frame of
fig.~\ref{fig:0.01-w-energy-frac}, which shows the three contributions
of eq.(\ref{math:friedman}) to the total energy density, averaged over
a couple of oscillations of the background field. We see that the
gradient term remains negligible until just before the instance of
collapse, but then very quickly becomes sizable. Initially, while the
field inhomogeneities are small, the kinetic and potential energies
oscillate out of phase with each other. In the linear approximation,
i.e. for harmonic oscillations, their average values should be the
same. Instead we see that initially the potential energy is about
$50\%$ larger, while after the collapse of the background field the
kinetic energy gives the biggest contribution.

The gray dotted line in this figure is the prediction of the Virial
theorem for the time averaged kinetic energy fraction:
\begin{equation} \label{math:virial}
  \rho_{K, \text{virial}} = \expval{\rho_G}
  + d \left( 2 \phi_0^2 \expval{\phi^2} - 4 \phi_0 \expval{\phi^3}
  + 2 \expval{\phi^4} \right)\,.
\end{equation}
A heuristic derivation can be found in appendix
\ref{sec:app-virial}. During slow--roll inflation, $\rho_G = 0$ and
$\phi \simeq \phi_0$. The virial theorem then predicts a very small
$\rho_K$, which is indeed the case. In fact, the actual $\rho_K$
closely tracks the prediction of the virial theorem during the entire
simulation, including the time when the classical field collapses. We
will come back to the issue of virialization in sec.~\ref{subsec:osci}.

The right frame of fig.~\ref{fig:0.01-w-energy-frac} shows the
equation of state parameter $\omega$, computed from
\begin{equation} \label{math:omega}
  \omega = \frac{p}{\rho}
  = \frac{\expval{\rho_K}- \expval{\rho_G}/3 - \expval{\rho_V}}
  { \expval{\rho_K} + \expval{\rho_G} + \expval{\rho_V} }\,.
\end{equation}
Of course, $\omega \simeq -1$ during inflation. We see that its time
average remains negative as long as the backreaction is small,
although $\omega > -1/3$ implies that the expansion of the universe is
slowing down while the background field oscillates coherently. When
the background field collapses, $\omega$ becomes positive but quite
small, $\omega \lesssim 0.1$; it very slowly decreases later on. Hence
the assumption $\omega = 0$ frequently made in analyses of reheating
if the inflaton potential is quadratic near the origin is not a bad
approximation, but might not be good enough when trying to precisely
relate the reheat temperature and the number of e--folds of inflation
after CMB scales left the horizon, usually called $N_{\rm CMB}$.

\subsection{Hartree approximation}

\begin{figure}[ht]
\centering
\includegraphics[width=0.85\textwidth]
{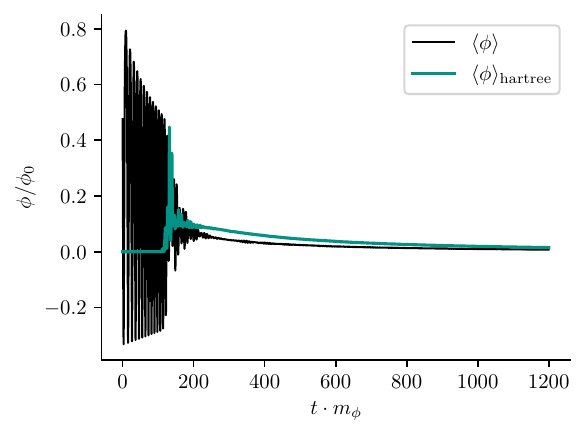}
\caption{The black curve again shows the evolution of the background field
  as computed from our lattice simulation, as in fig.~\ref{fig:0.01-mean}.
  The blue curve shows the value of the background field $\expval{\phi}$ that
  minimizes the Hartree potential (\ref{math:Hartree}).}
\label{fig:0.01-hartree}
\end{figure}

In this subsection we describe a first attempt to analyze the
backreaction of the fluctuations onto the mean field
$\expval{\phi}$. To that end we decompose the scalar field into the
homogeneous part and the perturbations:
\begin{equation} \label{math:decompose}
\phi(x, t) = \expval{\phi}(t) + \delta \phi(x, t)\, ,
\end{equation}
where $\expval{\delta \phi} = 0$ by definition. We insert this ansatz
into the inflaton potential (\ref{math:V-sfpi}), and keep terms that
are at least linear in $\expval{\phi}$ (the constant term doesn't
influence the evolution of the background field). Finally, we employ
the Hartree approximation, i.e. we average over the volume, which
removes terms linear in $\delta\phi$. This yields:\footnote{We neglect
  terms $\propto \beta$, which are only important if $\expval{\phi}$
  is very close to $\phi_0$.}
\begin{equation} \label{math:Hartree}
  V_{\rm Hartree} =  d \left[ \expval{\phi}^4 - \frac{8}{3} \phi_0
    \expval{\phi}^3 + \expval{\phi}^2 \left( 2 \phi_0^2 + 6
      \expval{\delta\phi^2} \right) + \expval{\phi} \left( 4
      \expval{\delta\phi^3} - 8 \phi_0 \expval{\delta\phi^2} \right) \right]\,.
\end{equation}
The result depends on the variance $\expval{\delta\phi^2}$ and
skewness $\expval{\delta\phi^3}$ that we had shown in
fig.~\ref{fig:0.01-delta-phi-2-3}. There is a positive contribution to
the coefficient of $\expval{\phi}^2$ which will increase the mass of
the background field; however, the left frame of
fig.~\ref{fig:0.01-delta-phi-2-3} shows that this effect is
significant only for $150 \lesssim t \cdot m_\phi \lesssim 200$. In contrast,
the term linear in $\expval{\phi}$ is new. Fig.~\ref{fig:0.01-delta-phi-2-3}
shows that the second, negative, contribution usually dominates. It
moves the minimum of the potential away from zero, to a field value
$\expval{\phi}_{\rm Hartree} \lesssim 2 \expval{\delta\phi^2} / \phi_0$.

The result is shown in fig.~\ref{fig:0.01-hartree}. Of course,
$\expval{\phi}_{\rm Hartree}$ does not oscillate. However, it allows
to understand why $\expval{\phi}$ increases again slightly just after
its initial collapse, and describes its long--term behavior
semi--quantitatively. Since we had to take $\expval{\delta\phi^2}$ and
$\expval{\delta\phi^3}$ from our simulation in order to compute
$\expval{\phi}_{\rm Hartree}$, this does not allow ab initio
understanding of the behavior of $\expval{\phi}(t)$ during and after
the collapse.  However, it is a useful consistency check on our
simulation.

\subsection{Field fluctuation spectrum}

Next we consider the Fourier modes of the field and density
perturbations. Although the power spectrum cannot capture all the
properties of the lattice results (e.g. there is definitely
non-Gaussianity present), it is still worth investigating. In the
following, the power spectrum is defined via the two--point
correlation function in Fourier space \cite{figueroaCosmoLattice2021}:
\begin{equation} \label{math:power}
  \expval{f(\vec{k}) f(\vec{k}') } = (2\pi)^3  \frac {2\pi^2} {k^3}
  \mathcal{P}_f(k) \delta^{(3)} \left(\vec{k} - \vec{k}' \right)\,.
\end{equation}
Here $f$ can be any scalar quantity, e.g. the field $\phi$
itself or an energy density, defined in Fourier space. Note that
rotational symmetry has been assumed, hence the power only depends on
the absolute value $k = |\vec{k}|$. The power can be used to directly
compute the second moment of $f$ in coordinate space:
\begin{equation} \label{math:power2}
  \expval{f^2} = \int \dd{\log k} \mathcal{P}_f(k)\,.
\end{equation}

\begin{figure}[ht]
\centering
\includegraphics[width=0.5\textwidth]
{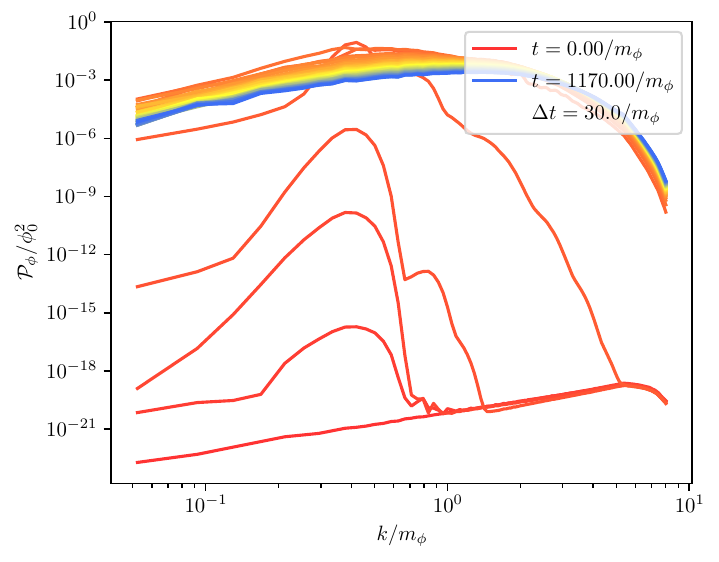}%
\includegraphics[width=0.5\textwidth]
{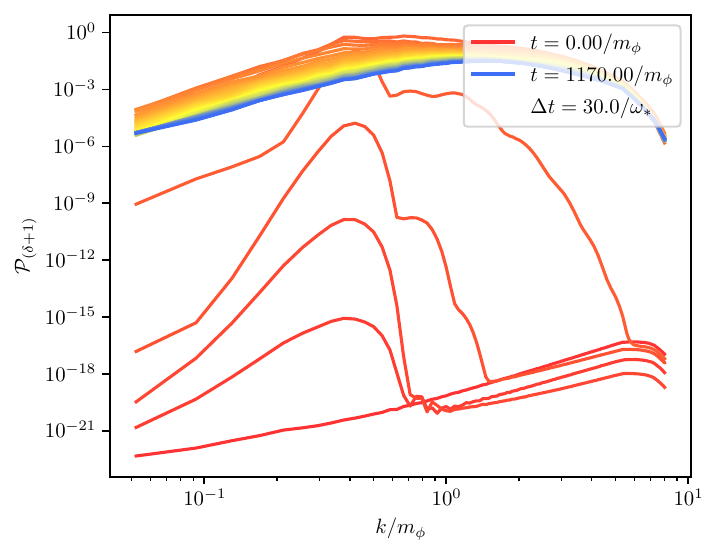}
\caption{The power spectrum of the inflaton field (left) and of the
  relative energy density $\rho(x)/\expval{\rho} = \delta(x) + 1$
  (right) as function of the comoving momentum on a log--log
  scale. The color of the curve indicates when the spectrum is taken;
  from early to late time: red, yellow and blue.}
\label{fig:0.01-fluc-spec}
\end{figure}

The power spectrum of inflaton field fluctuations is shown in the left
frame of fig.~\ref{fig:0.01-fluc-spec}. We see that only modes with
$k \lesssim m_\phi$ show rapid (exponential) growth at the beginning
of the simulation (red curves); this agrees with our previous linear
analysis in sec.~\ref{sec:floquet}. As expected, modes with
$k\sim 0.3 m_\phi$ grow fastest, see fig.~\ref{fig:floquet-max}.  In
between the first two curves, i.e. in the first $t=30\ m_\phi^{-1}$ of
the simulation, this mode grows by roughly $5$ orders of magnitude. On
the other hand, the Floquet analysis (see e.g. fig.~\ref{fig:floquet})
predicts
\begin{equation} \label{math:floq-power}
\mathcal{P}_\phi (k) \propto \exp(2 \realp{\mu_k} \Delta t)\,,
\end{equation}
with $\realp{\mu_k} \simeq 0.26 m_\phi$ for $k = 0.3 m_\phi$, or a growth
factor of $4 \cdot 10^6$ in this case. It is not surprising that the
Floquet analysis over--estimates the growth somewhat, since the full
lattice simulation takes into account the energy loss of the
background field due to Hubble friction as well as the redshift
effect; both effects tend to reduce the amount of
amplification.\footnote{In addition, there is a subtlety in the
  Floquet analysis, related to the periodic functions $A(t)$ and
  $B(t)$ in eq.(\ref{floquet-solution-general}). If one measures
  the growth rate in a time period which is not an integer multiple of
  the period, then the time dependence of these periodic functions
  influence the result.}

At later times modes with larger $k$ also begin to grow rapidly, as
can be seen from the fourth and fifth red curves in
fig.~\ref{fig:0.01-fluc-spec}. Recall that the condition for tachyonic
instability is $\omega_k^2 = k^2/a^2 + \tilde{m}^2 < 0$, so the
excitation of modes with higher comoving momentum is partly due to the
redshift.

At $t \approx 120 \ m_\phi^{-1}$, the spectrum reaches a peak value
$\lesssim 10^{-1} / \phi_0^2 $, which is comparable to the square of
the background field value shown in fig.~\ref{fig:0.01-mean}. The
brief time window afterwards when the excited modes get more evenly
distributed across all (available) momenta can be understood as
re--scattering \cite{dufauxPreheatingTrilinearInteractions2006}. Then
the spectrum slowly moves to the UV. Recall, however, that we plot the
spectrum vs. the co--moving momentum $k$, the physical momentum $q$
being smaller by a factor $a(t)$; recall that we set $a = 1$ at the
beginning of our simulation. Physical momenta redshift by a factor
$\sim 2.5$ in the course of our simulation. The peak of the power
spectrum at $k \sim 2 m_\phi$ at the end of our simulation (blue curve
in fig.~\ref{fig:0.01-fluc-spec}) therefore corresponds to
$q \sim 0.8 m_\phi$. Nevertheless this can be regarded as first
indication that quite small structures appear; we will see shortly
that this is indeed the case.

The right frame of fig.~\ref{fig:0.01-fluc-spec} shows the relative
energy density power spectrum.  It is again calculated from
eq.(\ref{math:power}), with $f$ now being the density contrast
\begin{equation} \label{math:rho-contrast}
  \delta (x) = \frac {\rho(x) - \expval{\rho}} {\expval{\rho}}
  = \frac {\rho(x)} {\expval{\rho}} - 1\, .
\end{equation}
Qualitatively the curves are similar to those in the left
frame. Quantitatively the modes with $k \gtrsim m_\phi$ grow a little
faster. The energy density gets a contribution from the kinetic
energy, which is naturally associated with somewhat larger values of
$k$.

\subsection{Oscillons}
\label{subsec:osci}

Clearly our system shows quite non--trivial dynamics, due to the
inflaton self interactons. We therefore expect distributions to be
quite non--Gaussian. Moreover, we saw at the end of the previous
subsection an indication for the formation or relatively small
structures. The latter can be investigated most directly by looking
at the lattice itself. To that end, we take ``snapshots'' of the
distribution of the (total) energy density across the lattice. From
it, other quantities, for example probability distribution functions,
can be extracted.

\begin{figure}[ht]
\centering
\begin{subfigure}[b]{0.49\textwidth}
  \includegraphics[width=\textwidth]
  {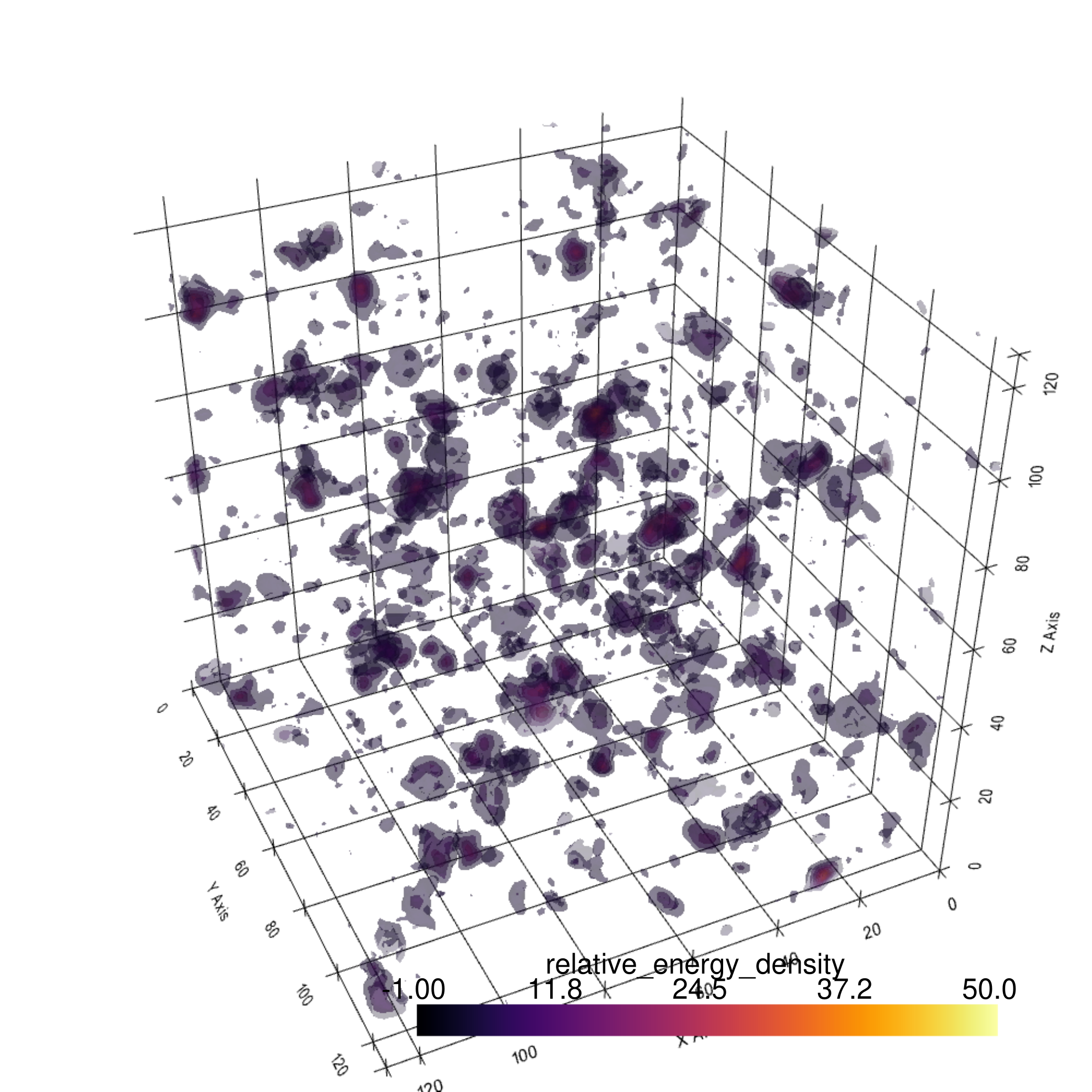}%
\caption{$t = 360 \ m_\phi^{-1}$}
\label{fig:0.01-snaps-1}
\end{subfigure}
\begin{subfigure}[b]{0.49\textwidth}
  \includegraphics[width=\textwidth]
  {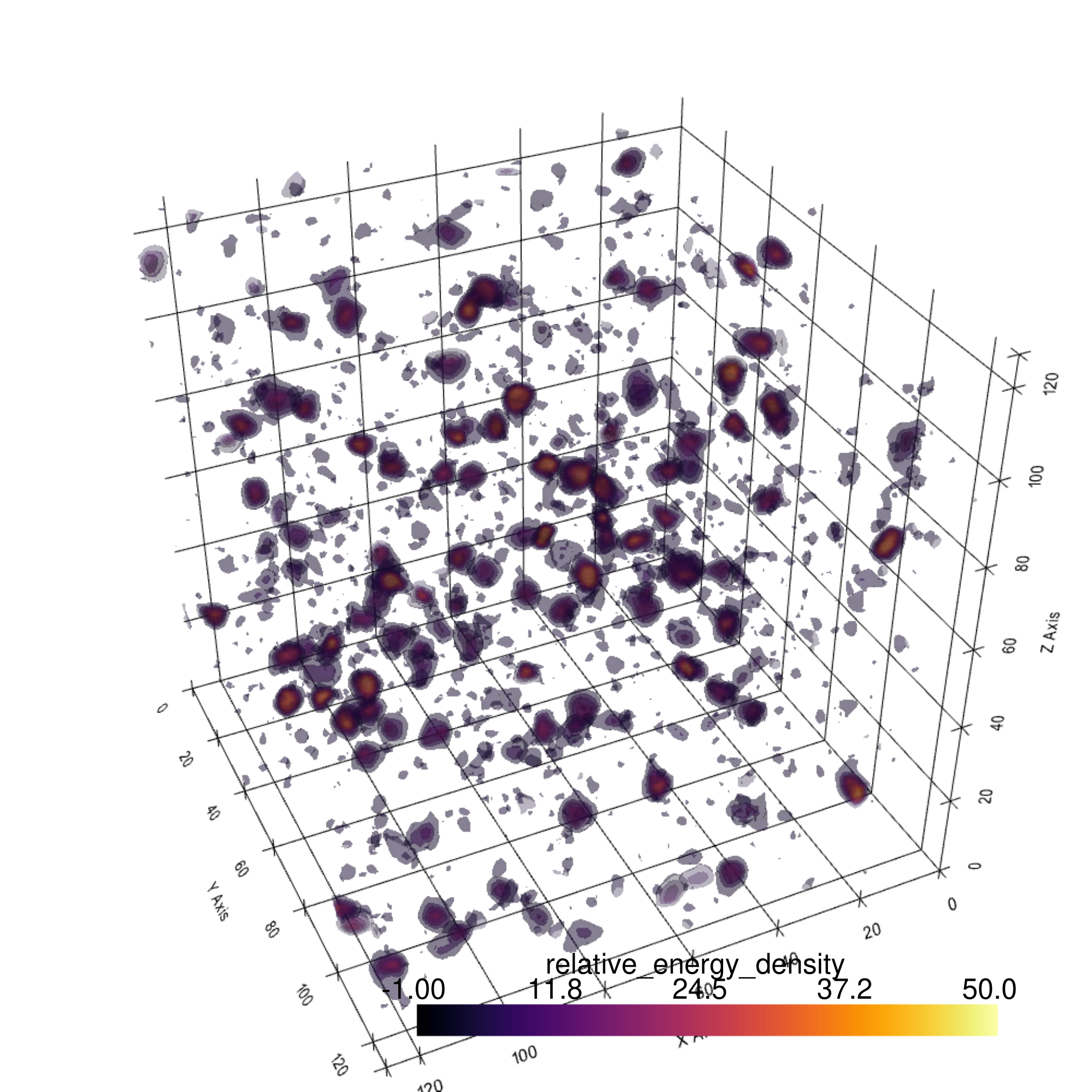}
\caption{$t = 960 \ m_\phi^{-1}$}
\label{fig:0.01-snaps-2}
\end{subfigure}
\caption{Isosurfaces of relative energy densities
  $\rho(x) / \expval{\rho}$ predicted by the lattice simulation at two
  different times. The axes denote comoving spatial coordinates in
  terms of the lattice spacing $\delta x$; in physical units,
  $\delta x \simeq 0.6/m_\phi$. For better visualization, only $1/8$th
  of the volume of the whole lattice is shown here.}
\label{fig:0.01-snaps}
\end{figure}

Figure \ref{fig:0.01-snaps} shows two such snapshots, taken after the
collapse of the initial, homogeneous field but before the
fragmentation process completely settled down. Already in the left
frame, taken at $t = 360/m_\phi$, small quasi spherical lumps are
formed; some of them have central relative energy density
$\rho(x) / \expval{\rho} \gtrsim 20$. Note that although these
objects are highly localized, they are not stationary; some tend to
move around. Indeed, there is nothing to forbid them from having a
velocity.

By $t = 960/m_\phi$ (right frame) these lumps have shrunk further; in
some cases the central overdensity now exceeds $30$. At this later
time the typical size of these lumps is about $5$ lattice units; from
eqs.(\ref{math:lattice-var}) and (\ref{math:lattice-spacing}), and the
values given in table \ref{tab:phi0-0.01} in Appendix \ref{sec:app-lat-para}, 
we see that
$\delta x \simeq 0.6/m_\phi$ in physical (albeit comoving) units. A
size of $3/m_\phi$ corresponds to
$k \simeq 2\pi \cdot (m_\phi/3) \simeq 2 m_\phi$. We saw in
fig.~\ref{fig:0.01-fluc-spec} that the late time density contrast
power spectrum indeed peaks at about this value of $k$. This is just a
qualitative consistency check; we will analyze the size distribution
of these overdense regions more quantitatively towards the end of this
subsection.

These objects can be identified as oscillons, which are
non--topological solitons with rapidly oscillating core
\cite{bogolyubskyi.lPulsedSolitonLifetime1976,
  gleiserPseudostableBubbles1994,
  copelandOscillonsResonantConfigurations1995}. Unlike its close
relative, the $Q-$ball of complex scalar field, an oscillon is not
associated with an exactly conserved charge and thus not infinitely
stable (by Derrick's theorem
\cite{g.h.derrickCommentsNonlinearWave1964}).

Nevertheless, under certain conditions, it has been known for some
time that a scalar field can fragment into such quasi--stable
configuration \cite{copelandOscillonsResonantConfigurations1995}. Of
course, this requires nontrivial self--interactions of this field; in
particular, the potential needs to be shallower than quadratic over an
extended range of field values
\cite{aminFlattopOscillonsExpanding2010}.\footnote{To see this in a
  crude way, consider a localized field configuration in one
  dimension: $\phi(x, t) = \Phi (x) \cos(\omega t)$. The equation of
  motion implies $ -\omega^2 \Phi - \partial_x^2 \Phi + V'(\Phi) = 0$.
  Away from the center of the configuration, one can take
  $V'(\Phi) \sim m^2 \Phi$, as the non--linearity should not be
  relevant. In order to have a localized configuration one needs to
  have positive curvature, $\partial_x^2 \Phi > 0$, near its boundary;
  this requires $\omega^2 < m^2$. In the center, however, we expect a
  maximum, i.e.  $\partial_x^2 \Phi < 0$. In order to achieve this
  while $\omega^2 < m^2$, one must have $V'(\Phi) < m^2 \Phi$.} Our
potential, given in eq.\eqref{math:V-sfpi}, clearly satisfies this
condition for $\phi > 0$, where the cubic term is negative; indeed, as
we emphasized in sec.~\ref{sec:model}, the potential has negative
curvature for $1/3 < \phi/\phi_0 < 1$, see eq.(\ref{range}). Note,
however, that our potential is {\em steeper} than quadratic for
$\phi < 0$. This is not the first time that oscillons have been observed
in a highly asymmetric potential, see 
\cite{mahbubOscillonFormationPreheating2023}.

Earlier studies using symmetric potentials found that oscillons can be
surprisingly stable, with lifetime up to $10^8 m_\phi^{-1}$. This is
well beyond the end of our simulation. In fact, since oscillons have
roughly constant physical size, our lattice becomes too coarse to
simulate them at $t \gg 10^3/m_\phi$. However, we see no reason why
our oscillons shouldn't decay eventually. In the known examples they
eventually transition to outgoing waves
\cite{zhangClassicalDecayRates2020, mukaidaLongevityIballOscillon2017,
  copelandOscillonsResonantConfigurations1995,
  levkovEffectiveFieldTheory2022}. Recall from our discussion at the
  end of sec.~\ref{sec:lattice-setup} that in our model perturbative decay of the inflaton
only happens at $t \geq 10^{14}/m_\phi$. It seems safe to assume that all
oscillons will have disappeared well before that time.

\begin{figure}[ht]
\centering
\includegraphics[width=0.95\textwidth]
{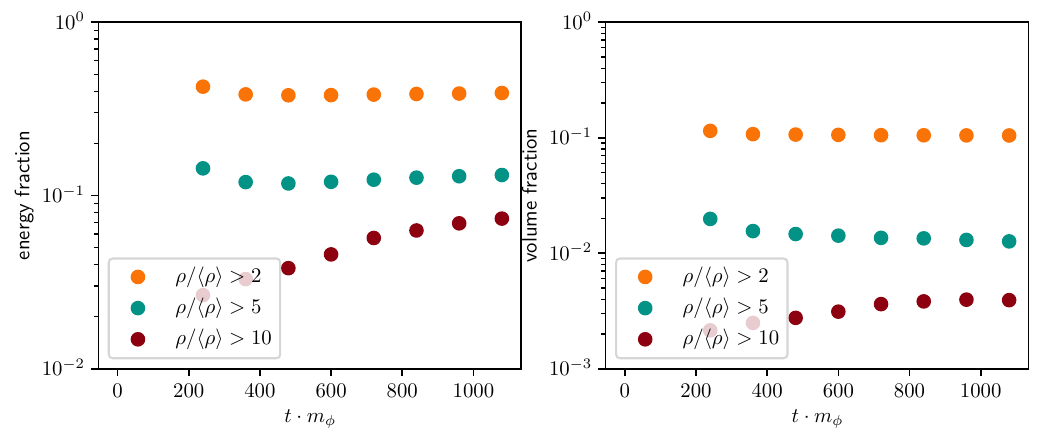}
\caption{Fraction of energy (left) and volume (right) of regions where
  the energy density exceeds the overall average by at least a factor
  of $2$ (orange), $5$ (blue) and $10$ (dark red).}
\label{fig:osci_frac}
\end{figure}

In order to understand the impact of oscillon formation more
quantitatively, we inspect which fraction of the energy these compact
objects contribute, and how much space they occupy. Results for
$t \geq 200 \ m_\phi^{-1}$ are shown in fig.~\ref{fig:osci_frac}. We
produced these results by simply scanning over all the points in the
lattice; a given point contributes (e.g.)  to the orange curves if the
local density is at least twice the global average. Prior to the
collapse of the background field, the inhomogeneities (resulting from
quantum fluctuations) are very small. Regions with overdensity of more
than $100\%$ (orange) first appear at $t \simeq 20\ m_\phi^{-1}$, but
fill only $\sim 3\cdot10^{-6}$ of the total volume. At
$t = 40 \ m_\phi^{-1}$, still well before the times covered in
fig.~\ref{fig:osci_frac}, nearly $5\%$ of points have more than
$100\%$ overdensity, and about $0.1\%$ of points have density more
than $10$ times the average value, shown in dark red in the
figure. Much later, at $t \simeq 800\ m_\phi^{-1}$, the system enters
a quasi--stable configuration, where $\sim 30 \%$ of the energy of the
whole lattice is stored in localized regions with
$\delta(x) = \rho/\expval{\rho} - 1 > 1$; these regions only occupy
$\sim 10 \%$ of the volume.

The results of fig.~\ref{fig:osci_frac} only show the total
contribution of all overdense regions. Ideally we would want to track
individual oscillons. Unfortunately, this is not so easy, since they
are quite small but move around fairly quickly. In order to at least
partially tackle this problem, we employ Density--Based Spatial
Clustering of Applications with Noise (DBSCAN) from \verb|sklearn|
\cite{scikit-learn}. We only consider oscillons that have
$\rho/\expval{\rho} > 10$ somewhere in their volume. We then label all
nearby points where $\rho/\expval{\rho} > 5$. DBSCAN is used to group
these points into discrete clusters. This gives at least a rough idea
of the energy contained in, and size of, individual oscillons. One
advantage of DBSCAN is the ability to remove anomalies, so that some
small non--oscillon fluctuations can be excluded. Of course, the
criterion for such anomalies must be given by hand.

\begin{figure}[ht]
\centering
\includegraphics[width=0.8\textwidth]
{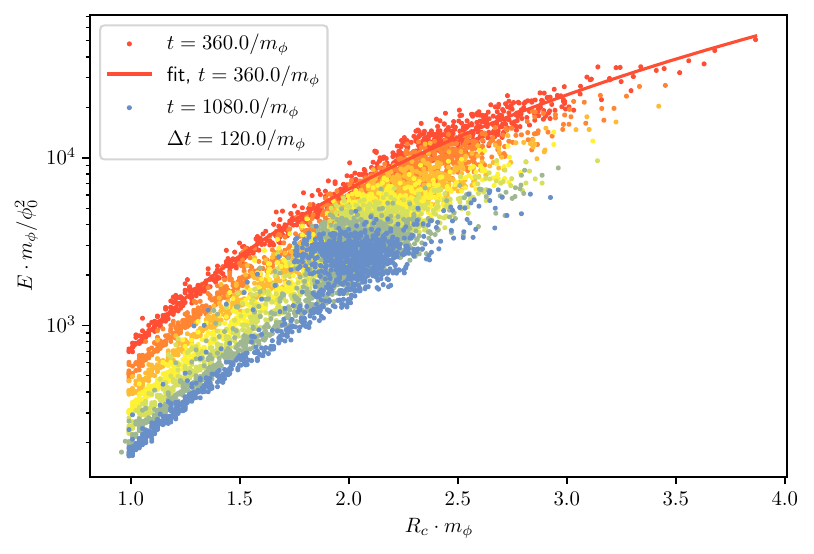}
\caption{Energy and (comoving) radius $R_c$ of oscillons taken at
  different times. Oscillons are identified and isolated using the
  DBSCAN algorithm. The criterion for oscillons used here is a high
  overdensity, $\rho/\expval{\rho} > 10$, and we consider nearby
  points with $\rho/\expval{\rho} > 5$ to belong to the same oscillon.
  The choice of other DBSCAN parameters doesn't affect the result much; 
  here we choose $\text{eps}=3$ and $\text{min-samples}=20$.}
\label{fig:osci-M-R}
\end{figure}

The energy and (comoving) radius of each oscillon are computed in the
following way:
\begin{subequations} \label{math:osc}
\begin{align}
E &= \sum_{\text{cluster}} (\delta x)^3 \rho\,,  \\
R_c & = \left( \frac{3}{4\pi} \sum_{\text{cluster}} 1 \right) ^{1/3} \delta x\,;
\end{align}
\end{subequations}
in the second equation we have implicitly assumed that oscillons are
spherically symmetric. In both equations the sum is taken over all the
points in one cluster from DBSCAN. At any given time (with
$t \geq 360 \ m_\phi^{-1}$) there will be many such clusters in our lattice. Their
energies are plotted versus their radii in fig.~\ref{fig:osci-M-R}.

We see that initially, i.e. at $t = 360 m^{-1}_\phi$, the algorithm
finds quite a broad spectrum of cluster sizes and energies. These
results can be described quite well by a power law,
$E \propto R_c^{\num{3.19}}$. The exponent only slightly deviates from
$3$, indicating that the configurations have nearly constant average
energy density. Recall, however, that at this time the system hasn't
settled into a near--equilibrium yet; among other things, we do not
expect the lumps to be strictly spherically symmetric.

The conventional wisdom is that non--linear localized structures don't
get much affected by the expansion of spacetime, similar to galaxies
in the expanding Universe. Since we lack the ability to track each
individual oscillon a definite statement cannot be made. However,
fig.~\ref{fig:osci-M-R} clearly shows that the energy of most
oscillons decreases quite significantly with time. This agrees with
ref.~\cite{antuschPropertiesOscillonsHilltop2019}, where a spherical
lattice simulation is used to trace one single oscillon for a long
time ($\order{10^5}$ oscillations), for a hilltop--shaped inflaton
potential. In their simulation a given oscillon can lose almost half
of its initial energy while the Universe expands by a factor of $5$.

Another intriguing feature of fig.~\ref{fig:osci-M-R} is that
gradually over--dense regions seem to move to
$R_c \approx 2 m_\phi^{-1}$ and
$E\cdot m_\phi / \phi_0^2 \approx 2000$; see the blue dots. Evidently
many over--dense lumps relax into this relatively stable configuration
after fragmentation; some of the lumps with lower radius and energy
might even be artifacts of the DBSCAN algorithm.

Ref.~\cite{copelandOscillonsResonantConfigurations1995} found that the
typical radius of an oscillon is $R_c \sim 3 m_\phi^{-1}$. As noted
above, in our case the lattice spacing can be related to the inflaton
(rest) mass via equation \eqref{math:lattice-spacing}. Here we find
roughly similar typical size of lumps. Quantitatively our oscillons
appear to be slightly larger, taking into account that the scale
factor in the FRW metric increases by a factor $\sim 2.5$ in the
course of our simulation. We note, however, that the oscillation
frequency, measured in units of $m_\phi$, also appears to be slightly
lower than in ref.~\cite{copelandOscillonsResonantConfigurations1995},
indicating that our effective mass (for the purposes of oscillon
physics) is slightly smaller than that in the (quite different)
potentials analyzed in this earlier reference.

We finally recall that our system remains virialized even while the
background field collapses, as noted in
section~\ref{subsec:average}. This is consistent with the observation
in ref.~\cite{copelandOscillonsResonantConfigurations1995} that
oscillons are virialized attractor field configurations.

\subsection{Dependence on $\phi_0$}
\label{sec:param}

It is time to check the effect of changing $\phi_0$, which is
basically the only free parameter of our model, on the preheating and
oscillon dynamics.

We first note that the system remains virialized at all times for all
values of $\phi_0$ we investigated, i.e. the kinetic energy averaged over
a couple of oscillations agrees very well with the prediction of the virial
theorem, as in fig.~\ref{fig:0.01-w-energy-frac}.

\begin{figure}[h!]
\centering
\includegraphics[width=0.6\textwidth]
{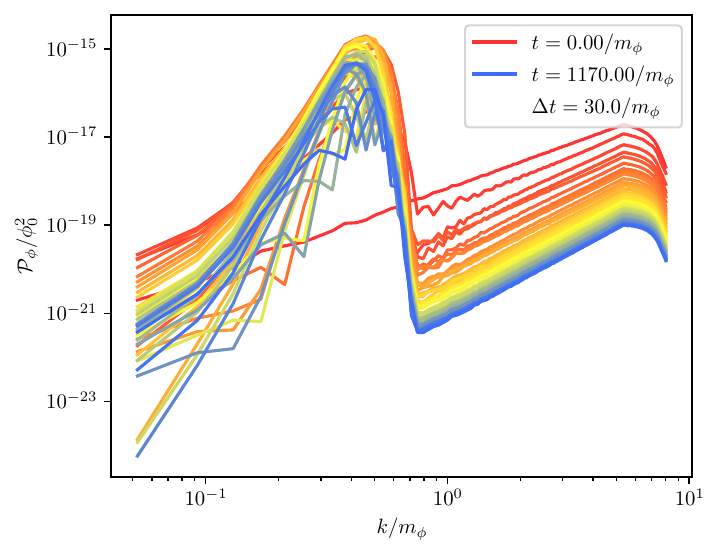}
\caption{Field fluctuation power spectrum for $\phi_0 = 0.1 \mpl$. }
\label{fig:0.1-power}
\end{figure}

As expected from results in section \ref{sec:floquet}, for large
$\phi_0$ the perturbations do not grow long enough to significantly
alter the background evolution. For example, we saw in
fig.~\ref{fig:background} that for $\phi_0 = 0.1 \ m_{\rm Pl}$ the
Hubble friction is so large that the background field only enters the
tachyonic region a few times. Hence the exponential growth of the
field fluctuation, whose spectrum initially again peaks around
$k = m_\phi/\sqrt{3}$, already ends at $t \simeq 30 \ m_\phi^{-1}$.
This is confirmed by fig.~\ref{fig:0.1-power}. Here we also see that
at later times the power in perturbations decreases again, due to
the continuing expansion of the universe.

The collapse of the background field occurs only for
$\phi_0 \lesssim 0.02 \mpl$. The dynamics here is already similar to
that of our benchmark point $\phi_0 = 0.01 \mpl$. For even lower
values of $\phi_0$ the dynamics remains qualitatively the same,
although some quantitative differences do occur. In the following we
show some results for $\phi_0 = 10^{-4} \ \mpl$.

\begin{figure}[h!]
\centering
\begin{subfigure}[b]{0.49\textwidth}
\begin{center}
  \includegraphics[width=\textwidth]
  {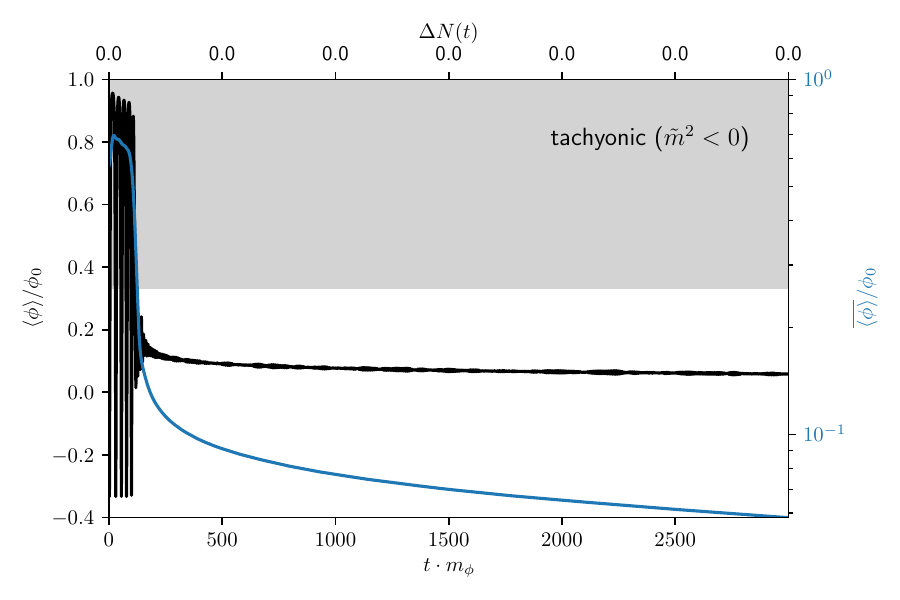}%
\end{center}
\end{subfigure}
\begin{subfigure}[b]{0.49\textwidth}
\begin{center}
  \includegraphics[width=0.9\textwidth]
  {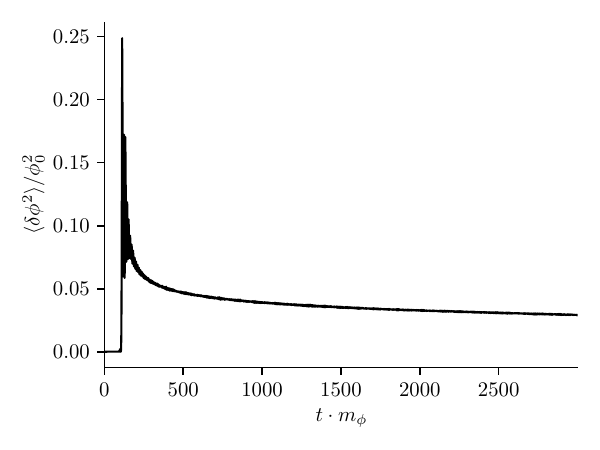}
\end{center}
\end{subfigure}
\caption{The spatial average of the inflaton field (left) and of its
  variance (right), for  $\phi_0 = 10^{-4}\ \mpl$.}
\label{fig:1e-4-delta-phi-2}
\end{figure}

The left frame of fig.~\ref{fig:1e-4-delta-phi-2} shows the evolution
of the background field; it should be compared with
fig.~\ref{fig:0.01-mean}.  The collapse of the background field again
happens at $t \simeq 100 \ m_\phi^{-1}$, since the increase of the
term $\propto \ln(\mpl/\phi_0)$ in eq.(\ref{math:back-react-time}) is
compensated by the slight increase in $\realp(\mu_k)$, see
fig.~\ref{fig:floquet-max}. However, recall from the discussion of
eq.(\ref{rho}) that $H/m_\phi \propto \phi_0$. Hence the expansion of
the universe during the simulation, shown by the upper horizontal axis
in the left frame of fig.~\ref{fig:1e-4-delta-phi-2} in terms of
$e$-folds, is negligible in this case, compared to an expansion by a
factor of $2.5$ for $\phi_0 = 0.01 \mpl$.

The right frame of fig.~\ref{fig:1e-4-delta-phi-2} shows the variance
of the inflaton field $\phi$, c.f. the left frame of
fig.~\ref{fig:0.01-delta-phi-2-3} which however uses a logarithmic
$y-$axis. The peak of the variance in units of $\phi_0^2$ is somewhat
higher for $\phi_0 = 10^{-4} \ \mpl$ ($0.25$ vs. $0.175$); more
importantly, in these units the variance stays near $0.05$ throughout
the simulation, whereas it quickly drops to about $0.02$ for
$\phi_0 = 10^{-2} \ \mpl$.

\begin{figure}[h!]
\centering
\begin{subfigure}[t]{0.5\textwidth}
\begin{center}
  \includegraphics[width=\textwidth]
  {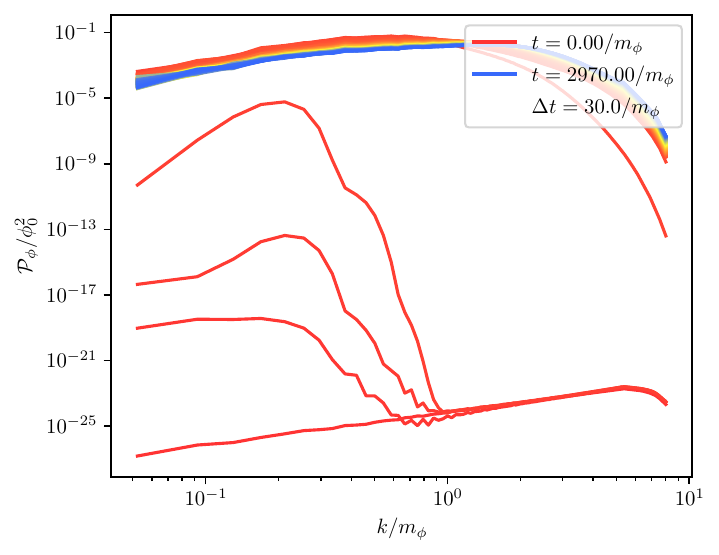}%
\end{center}
\caption{Power spectrum of fluctuations.}
\label{fig:1e-4-P}
\end{subfigure}%
\begin{subfigure}[t]{0.4\textwidth}
\begin{center}
  \includegraphics[width=\textwidth]
  {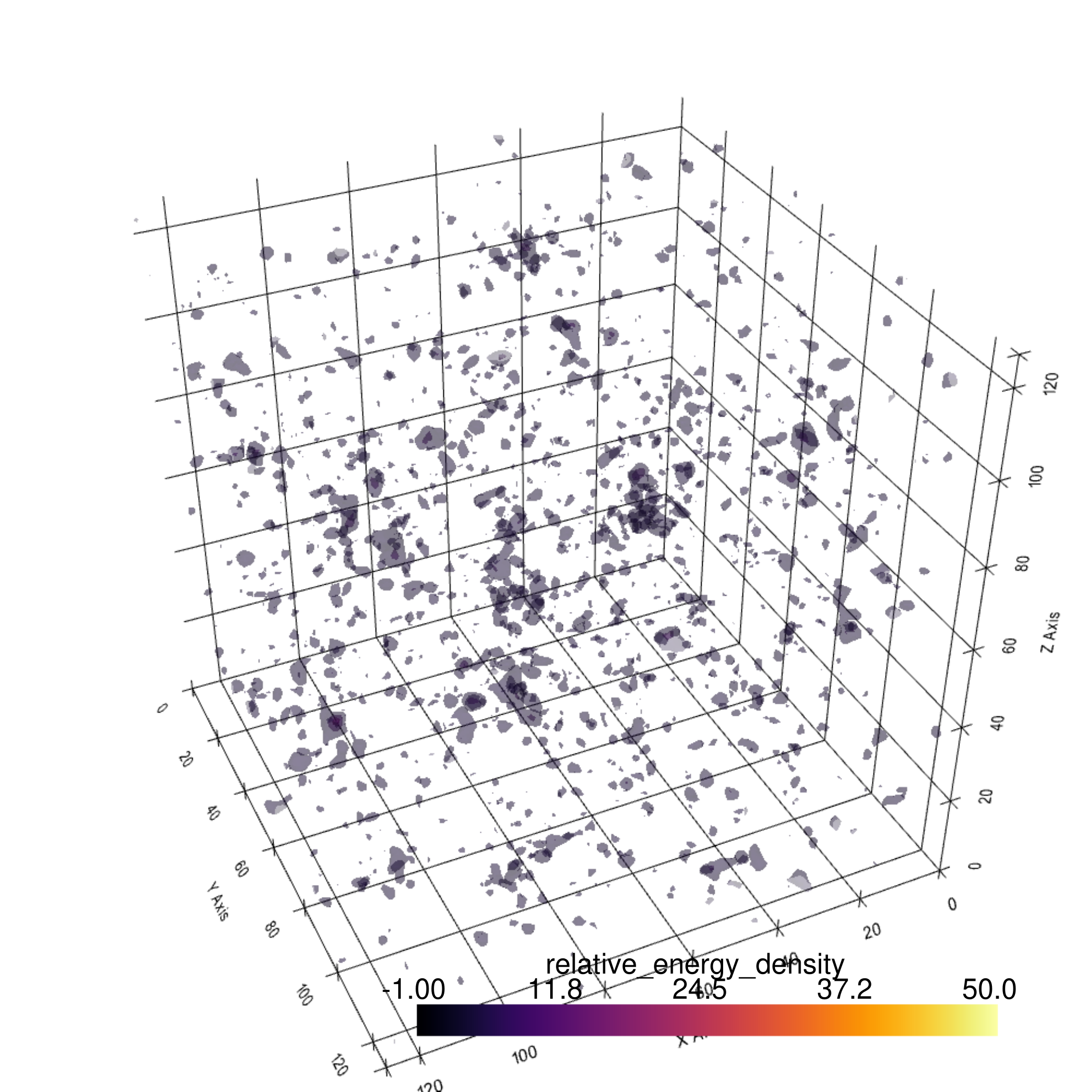}
\end{center}
\caption{Snapshot of the energy density contrast in $1/8$th of the
  lattice at $t = 1200 m_\phi^{-1}$.}
\label{fig:1e-4-delta}
\end{subfigure}
\caption{Power spectrum and snapshots for $\phi_0 = 10^{-4} \mpl$.}
\label{fig:1e-4-P-delta}
\end{figure}

The evolution of the power spectrum is depicted in
fig.~\ref{fig:1e-4-P}. We see that initially the amplified modes
always have comoving $k \lesssim m_\phi / \sqrt{3}$; due to the lack
of expansion the comoving $k$ remains very close to the physical one
in this case. The fastest growth occurs for $k \simeq 0.2 \ m_\phi$,
in agreement with the results of our Floquet analysis shown in
fig.~\ref{fig:floquet}. Modes with $k > m_\phi$ get excited only in
the nonlinear regime, once $\delta \phi_k \sim \expval{\phi}$. At late
times the peak value of $\mathcal{P}_\phi / \phi_0^2$ is slightly
higher than for $\phi_0 = 10^{-2} \ \mpl$.

The right frame of fig.~\ref{fig:1e-4-P-delta} shows a snapshot of the
total energy density at $t = 1200 \ m_\phi^{-1}$. Comparing with
fig.~\ref{fig:0.01-snaps} we see a lower typically density contrast of
$\lesssim 10$. The overdense regions (oscillons) here are typically
also smaller when measured in units of $m_\phi^{-1}$. Earlier
numerical studies of oscillons have found that their size and lifetime
can depend very sensitively on their initial core density
\cite{copelandOscillonsResonantConfigurations1995,
  aminFlattopOscillonsExpanding2010}. However, these simulations
typically start from a single oscillon with high initial density, and
are thus not directly comparable to our simulation where the initial
condition is very smooth.

\section{Observational consequences}
\label{sec:obs}

Our results so far show that for $\phi_0 \lesssim 0.02 \ \mpl$ the
inflaton field becomes very inhomogeneous and fragments into
oscillons. This happens at a time scale of $\sim 100 \ m_\phi^{-1}$,
well before the perturbative decay of the inflatons that reheats the
universe. It is important to ask whether this rather violent early
dynamics can have observable consequences.

We already saw in fig.~\ref{fig:0.01-w-energy-frac} that the
fragmented system has slightly positive equation of state parameter
$\omega$, while $\omega = 0$ for an inflaton field oscillating
coherently around a quadratic minimum. As a result, the Universe
expands slightly more slowly; recall that the scale factor
$a \propto t^{2/(3+3w)}$. This affects the number of e--folds before
the end of inflation when a given perturbation first exited the
horizon.  However, this also depends strongly on the reheat
temperature, which can vary over a considerable range in this model
\cite{dreesSmallFieldPolynomial2021}.

Recall also that earlier numerical studies found that oscillons live
for at most $\sim 10^8 m_\phi^{-1}$ before they decay by emitting
inflaton particles \cite{copelandOscillonsResonantConfigurations1995,
  zhangClassicalDecayRates2020,
  salmiRadiationRelaxationOscillons2012a}. If this result carries over
to our model, oscillons decay well before (free) inflatons do. This
indicates that the self resonance and oscillons have negligible
effects \cite{copelandOscillonsResonantConfigurations1995} on the
reheating dynamics. In the end, reheating will complete as usual, via
perturbative channel(s).

Nevertheless there are two possible messengers from the very early
universe, primordial black holes and gravitational waves. We will
discuss them in the subsequent subsections.

\subsection{Primordial black holes}

We saw in the previous section that large density contrasts can be
generated after the fragmentation. Here we analyze whether they could
seed black hole formation. If small black holes are produced, their
evaporation can impact reheating, and could even be its main
mechanism. At the same time, light black holes can also be a source of
relics, e.g.  dark matter particles.

We begin our argument by showing that the timescale for gravitational
collapse cannot be shorter than that for the fragmentation of the
inflaton field. For a static Universe, overdensities grow
exponentially by gravitational attraction, with typical timescale
(Jeans instability) \cite{rubakov2}
\begin{equation} \label{math:Jeans}
  t_J = \frac {1} {\sqrt{4\pi G \rho}} =
  \frac {\sqrt{2} \mpl} {\sqrt{{\rho}}} \frac {1} {\sqrt{\delta + 1}}
= \frac{1}{H} \frac {\sqrt{2}} {\sqrt{3 ( \delta + 1 )}}  \,.
\end{equation}
Here $\rho$ denotes the average energy density of a small (over--dense)
patch of the Universe and $\expval{\rho}$ is the average energy
density inside a Hubble horizon.\footnote{There should be an additional
  term containing the sound speed in the denominator. However, it
  increases the time scale of the gravitational instability.}
Therefore, if we assume that the density contrast cannot exceed unity
by more than a factor of $30$ or so, we have
\begin{equation} \label{math:times}
  t_J \gtrsim \frac{\sqrt{2}} {10 H} > \frac{\sqrt{2}} {10 H_I} \simeq
  \frac {\mpl} {m_\phi \phi_0}\,.
\end{equation}
Since nonlinear effects appear in the inflaton dynamics only if
$\phi_0 \lesssim 0.01 \ \mpl$, the inequality (\ref{math:times})
implies $t_J \gtrsim 100 \ m_\phi^{-1}$ for parameters that lead to
the fragmentation of the inflaton field. Hence gravitational collapse
cannot be faster than the collapse of the inflaton condensate into a
highly inhomogeneous configuration.

If PBHs are ever produced, then their masses will be
\begin{equation} \label{math:BH_mass}
M_{\text{BH}} = \frac{4\pi}{3} R^3 \rho_{\text{local}}\,,
\end{equation}
where $R$ is the typical extension of over--dense regions and
$\rho_{\text{local}}$ their energy density.  From our lattice
simulation,
\begin{equation} \label{math:osci_length}
R_o  \simeq  2 m_\phi^{-1}\,.
\end{equation}
Moreover, neglecting the expansion of space,
\begin{equation} \label{math:osci_rho}
\rho_{\text{local}} \sim 10 \expval{\rho}=30 \mpl^2 H_I^2\,.
\end{equation}
Inserting eqs.\eqref{math:osci_length} and \eqref{math:osci_rho} into
eq.\eqref{math:BH_mass} yields a mass of a single oscillon that is
independent of $\phi_0$:
\begin{equation} \label{math:pbh_mass}
  M_o \simeq 320 \pi  \frac{\mpl^2 H_I^2 }{m_\phi^3}
  \sim 2\cdot 10^3\; \si{\g}.
\end{equation}
No gravitational collapse would be needed if a single oscillon already
contained sufficient mass inside its radius to form a black hole. The
Schwarzschild radius of a (non--rotating) black hole is given by
\cite{stevenweinbergGravitationCosmologyPrinciples1972}
\begin{equation} \label{math:rs}
r_S = \frac{2}{8\pi} \frac{M}{\mpl^2}\, .
\end{equation}
Estimating the oscillon mass from eq.(\ref{math:pbh_mass}), the oscillon
radius of eq.(\ref{math:osci_length}) is smaller than the Schwarzschild
radius of eq.(\ref{math:rs}) if
\begin{equation}
\phi_0 / \mpl \gtrsim 1\,.
\end{equation}
Recall that oscillons only form if $\phi_0 \lesssim 0.01 \mpl$, in which
case the oscillon radius is at least two orders of magnitude larger than
the Schwarzschild radius.

Hence PBH formation would require the merger of many oscillons through
gravitational dynamics. This would have to occur while oscillons still
exist, i.e. at $t \lesssim 10^8 \ m_\phi^{-1}$. The maximal mass of a
black hole formed at time $t$ is given by the energy within one Hubble
radius, $1/H(t) = 3 t/2$ deep in a matter--dominated epoch
($w=0$).\footnote{In this simple estimate we ignore the deviation of
  $w$ from zero.} At such late time the dilution of the energy density
$\propto a^{-3}$ has to be taken into account. Recalling our
convention $a(t_0) = 1$ we have
\begin{equation} \label{math:a_of_t}
  a(t) = \left[ \frac{3}{2} H(t_0) (t-t_0) + 1 \right]^{\frac{2}{3(1+w)}}\,.
\end{equation}
For example, for $t-t_0 = 10^8 \ m_\phi^{-1}$ and taking $H(t_0) = H_I$,
we have
\begin{equation} \label{math:acube}
  a(t=10^8/m_\phi)^3 \simeq 10^{16} \left( \frac {\phi_0} {4\mpl}\right)^2\,.
\end{equation}
The mass inside a Hubble radius, $M_H(t) \simeq H(t)^{-3} \rho_I a(t)^{-3}$,
could then be estimated as
\begin{equation} \label{math:Hubble_mass}
  M_H(t=10^8/m_\phi) \sim 10^8 M_o \left( \frac {\mpl} {\phi_0} \right)^2\,.
\end{equation}
A primordial black hole would have to have a mass of a few times
$10^{15}$ g at least in order to not decay by Hawking radiation before
the present time \cite{CarrPBH2016}. Eqs.(\ref{math:Hubble_mass}) and
(\ref{math:pbh_mass}) show that this might conceivably happen for all
$\phi_0 \lesssim 0.01 \ \mpl$ where oscillons form. In contrast, since
the lifetime of a black hole scales with the third power of its mass,
a single oscillon somehow collapsing into a black hole would evaporate
after $\sim 10^{-19}$ seconds.

Of course, eq.(\ref{math:Hubble_mass}) only gives a crude upper bound
on the mass of any PBH that might conceivably form out of merging
oscillons. The mass inside a Hubble horizon (at appropriately chosen
time) is also used in the Press--Schechter formalism, which
automatically includes the gravitational collapse of over--dense
regions at later times. However, usually it is applied to
\textit{superhorizon, Gaussian} curvature perturbations; neither of
these two requirements is met here. There are some arguments that the
PS formalism might still apply in more general situations
\cite{lythFormingSubhorizonBlack2006a, youngPrimordialBlackHoles2013,
  suyamaBlackHoleProduction2006}, which however do not include
oscillons. It seems to us that the formation rate of PBHs in our
scenario can be estimated reliably only using methods of full
numerical relativity \cite{nazariOscillonCollapseBlack2021,
  kouOscillonPreheatingFull2020}. Some attempts have been made in
\cite{cotnerAnalyticDescriptionPrimordial2019,
  lozanovGravitationalPerturbationsOscillons2019,
  cotnerPrimordialBlackHoles2018,
  torres-lomasFormationSubhorizonBlack2014}.
Recently, a full-fledged numerical relativity simulation of oscillons
has been done; it seems unlikely for black hole to form in such scenarios \cite{Aurrekoetxea:2023jwd}.

\subsection{Gravitational waves}

The formation of black holes is an intrinsically non--perturbative
process.  Unfortunately we lack the computational resources to derive
quantitative results beyond the simple estimates presented in the
previous subsection. In contrast, the production of gravitational waves
can be treated perturbatively in many cases. Moreover, they do not decay,
nor are they diluted or otherwise significantly modified by the later
evolution of the universe, apart from the kinematic redshifting due
to the Hubble expansion. 

In this subsection we first briefly review the theory of gravitational
wave production from a scalar field in the (very) early universe. We
then compute the peak frequency of, and the energy density carried by,
primordial gravitational waves in our model.

\subsubsection{Theory}

This section reviews the basics of gravitational waves (at first
order), mostly based on
\cite{capriniCosmologicalBackgroundsGravitational2018}. Consider the
metric composed of the FRW background and a small perturbation in the
comoving frame:
\begin{equation}
	g_{\mu\nu} = a^2(\eta) (\eta_{\mu\nu} + h_{\mu\nu} )\,.
\end{equation}
We use the gauge freedom to set the temporal components as well as the
trace and ``divergence'' of $h$ to zero:
\begin{equation}
h_{\mu 0} = 0, \quad  h_{i}^{i} = 0,\quad \partial_i h_{ij} = 0.
\end{equation}
This is known as the transverse--traceless (TT) gauge. With this gauge
choice, we go from ten degrees of freedom to only two. The physical
results discussed below are gauge--independent to the order considered
\cite{capriniCosmologicalBackgroundsGravitational2018}.

One can decompose the metric perturbation into irreducible
representations of $\mbox{SO}(2)$. The symmetric transverse traceless
part $h_{ij}$ has helicity$-2$ and should be interpreted as
gravitational waves \cite{rubakov2}. The energy--momentum tensor can
also be decomposed into helicity sectors; the helicity$-2$ part is
$T_{ij}, i\neq j$. For an ideal fluid the off--diagonal entries of
$T_{ij}$ vanish, so that $h_{ij}$ is decoupled from the
energy--momentum tensor. But if one relaxes this condition, one
obtains \cite{capriniCosmologicalBackgroundsGravitational2018}
\begin{equation} \label{math:gw-h-EOM}
  \ddot{h}_{ij} + 3 H \dot{h}_{ij} - \frac{\nabla^2}{a^2} h_{ij}
  = \frac{2}{\mpl^2} \Pi_{ij}^{TT}\,;
\end{equation}
$\Pi_{ij}^{TT}$ is the transverse-traceless part of the anisotropic
tensor $\Pi_{ij}$, defined via
\begin{equation}
a^2 \Pi_{ij} = T_{ij} - p g_{ij}\,.
\end{equation}
For a single real scalar field $\phi$ one has
\cite{kouGravitationalWavesFully2021,
  capriniCosmologicalBackgroundsGravitational2018}
\begin{equation}
\Pi^{TT}_{ij} = (\partial_i \phi \partial_j \phi)^{TT}\,.
\end{equation}
Note that this term vanishes in a perfectly spherically symmetric
system, in accordance with Birkhoff's theorem
\cite{georgedavidbirkhoffRelativityModernPhysics1927}. The over--dense
regions created in the fragmentation of the inflaton background field
break this symmetry; in fact, as shown in the previous sections, even
a single such region is usually not perfectly spherically symmetric,
especially at early times. We therefore expect GWs to be produced when
the inflaton field fragments into oscillons. They will give rise to a
background of stochastic gravitational waves.

In order to quantify the magnitude of this contribution, one needs to
go the second order in metric perturbation. The effective
energy--momentum tensor of GWs is obtained by averaging the second
order Ricci tensor. Its $00$ component is
\begin{equation} \label{math:gw-h-rho}
\rho_{\gw} = T^{00}_\gw = \frac{\mpl^2 \expval{\dot{h}_{ij} \dot{h}^{ij}}}{4}\,,
\end{equation}
where the average is taken over length scales. In order to decide
whether this can be detected, one should consider the spectrum of GWs
rather than the integrated energy density:
\begin{equation} \label{math:gw-omega-def}
  \Omega_\gw (t) = \int \frac {1} {\rho_c(t)} \dv{\rho_\gw (k, t)}{\log k}
  \dd{\log k} := \int \Omega_\gw (k, t) \dd{\log k}\,.
\end{equation}
Here $\rho_c (t) = 3 \mpl^2 H^2(t)$ is the (time dependent) critical
energy density.

$\rho_\gw$ redshifts like radiation. As shown in
appendix~\ref{sec:app-gw-prop}, a gravitational wave produced with
comoving wave vector $k_\gw$ now has frequency
\cite{lozanovGravitationalPerturbationsOscillons2019}
\begin{equation} \label{math:gw-f0-og}
  f_0 = \frac{1}{2\pi} \frac {k_\gw} {a_p} \frac {a_p} {a_0}
  \simeq \frac {k_\gw} {a_p} \frac {1} {\rho_p^{1/4}} \cdot
  \num{4e10}  \si{\hertz}\, ,
\end{equation}
where the subscript $p$ denotes the time when gravitational waves were
produced whereas the subscript $0$ refers to the present time. The
scaled energy density $\Omega_\gw$ is constant during the radiation
dominated epoch, but becomes smaller in the matter and dark energy
dominated epochs, so that today,
\begin{equation} \label{math:gw-Omega0}
\Omega_{\gw, 0} \simeq \num{8e-6} \cdot \Omega_{\gw, p}\, .
\end{equation}
It should be noted that the approximation used in
appendix~\ref{sec:app-gw-prop} assumes a very short reheating period;
the results then don't depend on the reheating temperature. As shown
in appendix~\ref{sec:app-gw-reheating} this is not a bad approximation
for high reheat temperature $T_R$; however, for lower $T_R$ the frequency
and scaled energy density of primordial GWs might be further reduced by
a couple of orders of magnitude.

\subsubsection{Analytical Estimates}

Our model has basically only a single free parameter, $\phi_0$.  It
determines the mass of the inflaton, which in turn sets the scale for
the size of the overdense regions, as we saw above. We therefore write
$k_\gw = \tilde{k}_\gw m_\phi$, where the dimensionless quantity
$\tilde{k}_\gw$ should be of order unity if the typical wavelength of
GWs at production is comparable to the size of an oscillon.  Since
during the fragmentation of the inflaton field the universe expands by
less than one e--fold, we take
$\rho_p \simeq \rho (\phi = \phi_0) \simeq V(\phi_0)$. Then we obtain
from eq.\eqref{math:gw-f0-og}, setting $a_p = 1$:\footnote{Recall that
  we set the scale factor $a=1$ at the start of our simulation.}
\begin{equation} \label{math:gw-f0-sfpi}
  f_0 \simeq \tilde{k}_\gw \sqrt{\frac {\phi_0}{\mpl}} \cdot
  \SI{1.6e7}{\hertz}\,.
\end{equation}
Hence, smaller $\phi_0$ leads to lower frequency GW signals,
potentially in the sub $\si{\mega\hertz}$ range.

It is conjectured that at production the relative energy density of
GWs can be written as \cite{felderNonlinearInflatonFragmentation2007,
  dufauxTheoryNumericsGravitational2007}
\begin{equation}
\frac{\rho_{\gw, \tot}}{\rho_{\tot}} = \alpha \left( \frac{H}{k} \right) ^2\,,
\end{equation}
where the ``fudge factor'' $\alpha$ should be order of unity for a
sufficiently inhomogeneous medium emitting GWs. The current scaled GW
energy density can then be estimated as
\begin{equation} \label{math:gw-Omega0-sfpi}
  \Omega_{\gw, 0} h^2 \simeq \num{1e-7} \left( \frac{\phi_0}{\mpl} \right) ^2
  \frac {\alpha } {\tilde{k}_\gw^2}\,.
\end{equation}

\subsubsection{Numerical Results}

The evolution of the metric perturbation $h_{ij}$ is computed
numerically by discretizing eq.\eqref{math:gw-h-EOM}; the GW energy
density can then be obtained from eq.\eqref{math:gw-h-rho}. Finally,
today's GW frequency and energy density can be computed from
eqs.\eqref{math:gw-f0-og} and \eqref{math:gw-Omega0}, respectively
\cite{figueroaArtSimulatingEarly2021,
  baeza-ballesterosCosmoLatticeTechnicalNote}.

\begin{figure}[ht]
\centering
\includegraphics[width=0.7\textwidth]
{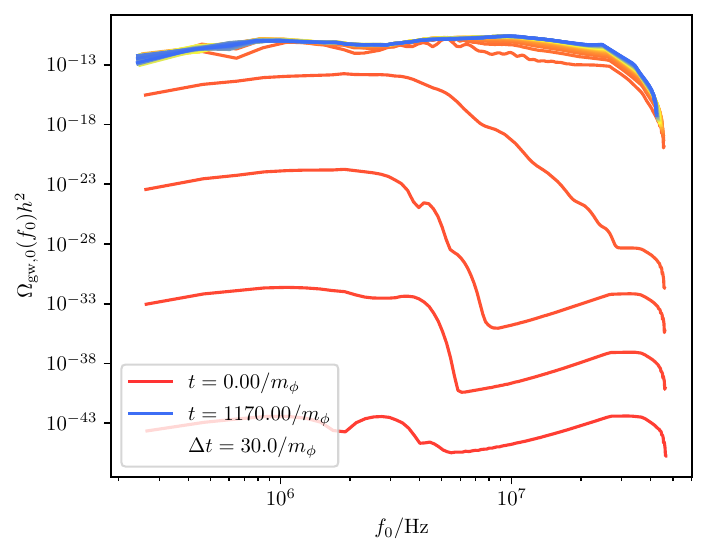}
\caption{Today's GW energy density per logarithmic frequency interval
  in terms of today's frequency for $\phi_0 = 0.01\ \mpl$. Today's reduced
  Hubble constant is taken to be $h=0.674$ \cite{planck6}. Each curve
  shows the contribution of GWs produced up to the time shown.}
\label{fig:gw-phi0-0.01}
\end{figure}

Results for $\phi_0 = 0.01\ \mpl$ are shown in
fig.~\ref{fig:gw-phi0-0.01}. We saw in fig.~\ref{fig:0.01-fluc-spec}
that initially field fluctuations with $k \lesssim m_\phi/2$ grow
exponentially due to the tachyonic instability. This gives rise to
gravitational waves with (today's) frequency
$\lesssim 5 \ \si{\mega\hertz}$. The broader, but less pronounced peak
in the GW spectrum at frequencies of tens of $\si{\mega\hertz}$ is
sourced later, during the epoch of linear growth of fluctuations and
their back--reaction on the inflaton condensate
\cite{zhouGravitationalWavesOscillon2013, antuschWhatCanWe2018}.  Once
oscillons are fully formed, they become quite spherical; as argued
above, a single oscillon will then not produce significant amounts of
GW any more, explaining the saturation of the curves at later times
\cite{lozanovGravitationalPerturbationsOscillons2019,
  dufauxPreheatingTrilinearInteractions2006}.\footnote{The motion of
  the various oscillons relative to each other does not end once
  oscillons are fully formed. However, the typical time scale for that
  is much longer than that for the very rapid field oscillations
  inside the oscillons -- hence their name, after all. Therefore the
  GWs produced in the encounters of separate oscillons have much lower
  energy density, and remains essentially invisible on the logarithmic
  scale of fig.~\ref{fig:gw-phi0-0.01}.}

\begin{figure}[ht]
\centering
\includegraphics[width=0.8\textwidth]{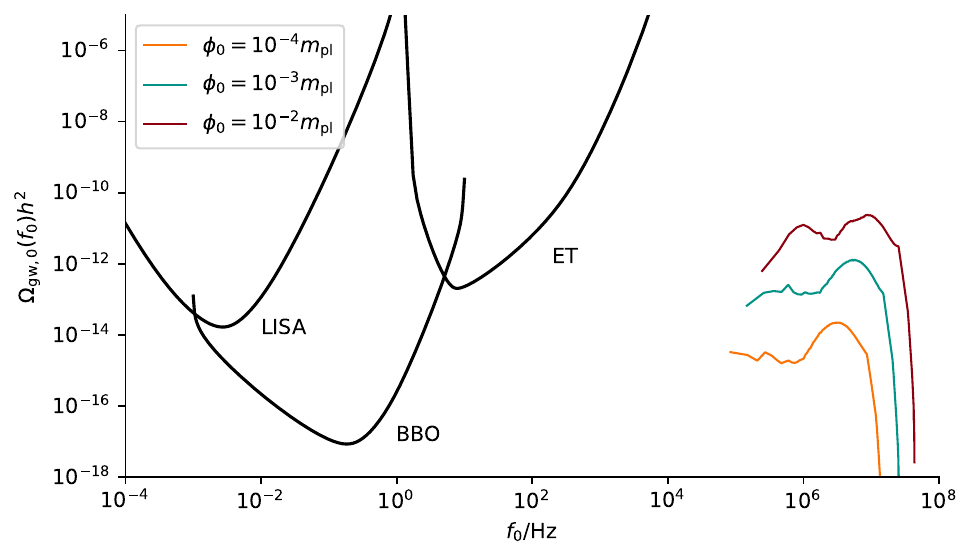}
\caption{The colorful curves show GW signals for three values of
  $\phi_0$; these correspond to the topmost curve shown in
  fig.~\ref{fig:gw-phi0-0.01}, i.e. they show the total of all GWs
  emitted during our simulation. For comparison, sensitivity curves
  for future experiments \cite{schmitzNewSensitivityCurves2021} are
  shown in black. }
\label{fig:gw-sens}
\end{figure}

In fig.~\ref{fig:gw-sens} we show the (final) GW energy density for
three values of $\phi_0$, along side with sensitivity curves of some
future experiments; the latter are taken from
ref.\cite{schmitzNewSensitivityCurves2021}. As indicated in
eq.(\ref{math:gw-f0-sfpi}), the peak of the GW spectrum decreases with
decreasing $\phi_0$. Numerically the scaling is slower than
$f_{\rm peak} \propto \sqrt{\phi_0}$. Moreover, as predicted by
eq.(\ref{math:gw-Omega0-sfpi}) $\Omega_\gw$ quickly decreases with
decreasing $\phi_0$. Unfortunately the GWs have peak frequency of
several $\si{\mega \hertz}$, out of reach for currently planned
next--generation GW detectors; some ideas for detecting such high
frequency GWs are discussed in
\cite{aggarwalChallengesOpportunitiesGravitational2021}. Note also
that the irreducible gravitational wave background from perturbations
of the metric during inflation, customarily parameterized by $r$, is
constrained by CMB experiments in the
$10^{-19} - 10^{-17} \si{\hertz}$ range
\cite{capriniCosmologicalBackgroundsGravitational2018}. Thus, the high
frequency gravitational waves could provide us the complementary
evidence of the dynamics of the inflaton field on the opposite side of
the spectrum.

Depending on the equation state of the system between fragmentation
and the end of thermalization, the curves shown in
fig.~\ref{fig:gw-sens} can move around somewhat. In particular, slow
thermalization, i.e. a low reheat temperature, implies extra
expansion, in which case the gravitation waves would have smaller
frequency (longer wavelength) and smaller energy density in today's
Universe, see appendix \ref{sec:app-gw-reheating}.

\subsubsection{Gravitational waves contribution to dark radiation}

Gravitational waves contribute to the radiation field as far as
the expansion history of the universe is concerned. Any measurement of
the background evolution can therefore in principle constrain the
total energy density carried by gravitational waves. This is usually
expressed by the effective number of neutrino generations
$N_{\rm eff}$ (after $e^+ e^-$ annihilation).

Processes in the early universe freeze out at temperature $T$ where
$\Gamma (T) \simeq H(T)$. In particular, the weak interactions between
neutron and proton freeze out at $T \sim 1\ \si{\mega\eV}$, the exact
temperature being sensitive to the expansion rate at that time. The
freeze--out temperature determines the neutron to proton ratio which
in turn determines the primordial ${}^4 \text{He}$ abundance, since
almost all neutrons end up in ${}^4 \text{He}$. Here we assume that
the photon to baryon ratio, which also affects BBN dynamics, is
determined from CMB data, which impose further constraints on
$N_{\rm eff}$.  Altogether, $N_{\rm eff} < 3.2$ at $95 \%$ confidence
level \cite{cyburtNewBBNLimits2005} corresponding to
$\Omega_{\gw, 0}h^2 < \num{1.12e-6}$
\cite{capriniCosmologicalBackgroundsGravitational2018}.\footnote{There
  are two conditions for this constraint to apply. First, the GWs must
  be produced before BBN. Second, the GWs must be sub--horizon at the
  time of BBN. Since the GWs considered here are produced after
  inflation, they remain inside the horizon until the present time.}

The spectra shown in fig.~\ref{fig:gw-sens} can be integrated via
eq.\eqref{math:gw-omega-def}. The resulting total scaled energy
density in and today's peak frequency of the gravitational waves
produced by the collapse of the homogeneous inflaton field are
collected in table~\ref{tab:label}, for four different values of
$\phi_0$.

\begin{table}[htpb]
\centering
\begin{tabular}{ccc}  \toprule
$\phi_0 / \mpl$ &  $f_{\max, 0}/\si{\hertz}$ &  $\Omega_{\gw, 0} h^2$\\ \midrule
\num{1e-2} & \num{ 8.8125e+06} & \num{ 2.0499653360059258e-11}\\
\num{5e-3} & \num{ 6.9170e+06} & \num{ 7.910478357677936e-12}\\
\num{1e-3} & \num{ 5.2492e+06} & \num{ 8.792978279089547e-13}\\
\num{1e-4} & \num{ 3.0015e+06} & \num{ 1.5687329234829918e-14}\\ \bottomrule
\end{tabular}
\caption{Today's peak frequency and integrated scaled energy density
  of GWs produced just after the end of inflation, for various values
  of $\phi_0$.}
\label{tab:label}
\end{table}

\begin{figure}[ht]
\centering
\includegraphics[width=0.7\textwidth]{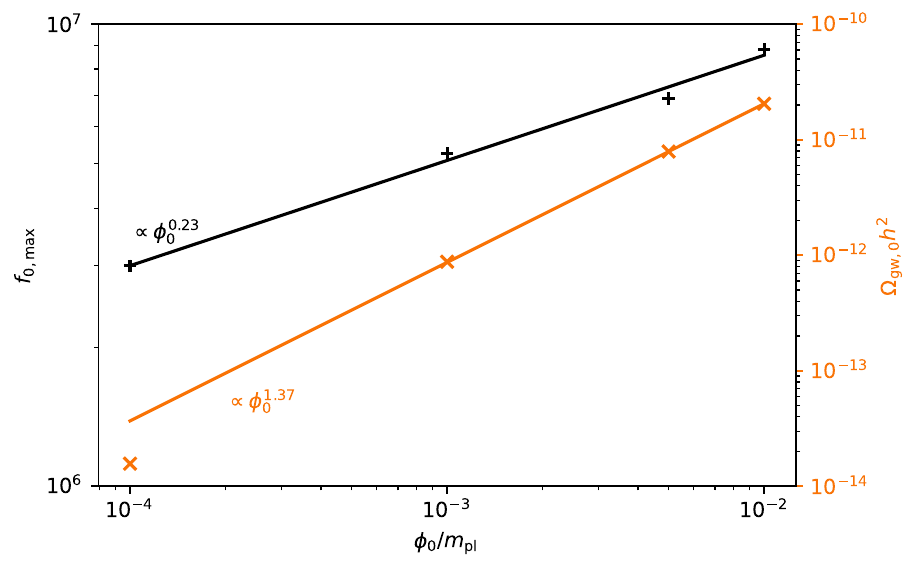}
\caption{Power law fits of today's peak frequency and integrated
  energy density of GWs produced just after inflation.}
\label{fig:gw_pl_fit}
\end{figure}

The scaling of $f_{0, \text{max}}$ and $\Omega_{\gw,0 }h^2$ with
$\phi_0$ are somewhat weaker than in the simple analytical estimates
of eqs.\eqref{math:gw-f0-sfpi} and \eqref{math:gw-Omega0-sfpi}. They
do nevertheless exhibit power law behavior, as shown in
fig.~\ref{fig:gw_pl_fit}. There is additional $\phi_0$ dependence
``hidden'' in $\tilde{k}_\gw$ and $\alpha$ which partly compensates
the decline of $f_{\max,0}$ and $\Omega_\gw$ with decreasing $\phi_0$.
On the other hand, we saw in sec.~\ref{sec:param} that for
$\phi_0 \gtrsim 10^{-2} \, \mpl$ the expansion of the universe during
the epoch of GW production is not negligible, hence the assumption
$\rho_p \simeq V(\phi_0)$ is not well warranted; this strengthens the
dependence of $f_{\max, 0}$ on $\phi_0$. In any case,
fig.~\ref{fig:gw_pl_fit} shows that $\Omega_\gw h^2$ is more than four
orders of magnitude below the present bound; its impact on the overall
evolution of the universe will therefore remain undetectable for the
foreseeable future.

\section{Conclusion}
\label{sec:conc}

In this work we investigated the post--inflationary non--perturbative
dynamics of the inflaton sector in a simple, renormalizable model with
a polynomial potential featuring a near inflection point at field
value $\phi = \phi_0$. As in (nearly) all successful models of small
field inflation, the second derivative of the inflaton potential is
negative during inflation. In our model the inflaton field re--enters
this tachyonic regime many times if the parameter
$\phi_0 \lesssim 10^{-1} \ \mpl$. As a result field fluctuations with
comoving wave vector $k \lesssim m_\phi/2$ grow exponentially; here
$m_\phi$ is the physical mass of inflaton particles. As long as the
back--reaction of the perturbations on the inflaton background can be
neglected the problem can be solved approximately using Floquet
analysis. The numerical results, in terms of Floquet exponents, tell
us how fast this growth is; see fig.~\ref{fig:floquet}. One can also
estimate the time of back--reaction, see
eq.(\ref{math:back-react-time}); at this time the background field
fragments, and the exponential growth of perturbations ends. Note that
even before this time the Floquet calculation involves several
approximations, hence the results cannot be fully trusted especially
when $\phi_0 / \mpl$ is relatively large. Analytically, this is a hard
problem, partly due to the nature of the potential since quadratic,
cubic and quartic terms are of comparable importance in the tachyonic
regime. An attempt has been made by fitting the effective mass as a
function of time. The height and position of the dominant peak as well
as the multiple bands in the Floquet map can be reproduced for small
$\phi_0$ with a simple box model, see fig.~\ref{fig:flo-ana}.

For large $\phi_0$ the Hubble friction quickly dampens the
oscillations of the background field, so that the exponential growth
of fluctuations soon comes to an end. On the other hand, for
$\phi_0 \lesssim 0.02\ \mpl$, the fluctuations become large enough to
significantly alter the background field evolution, i.e. the
back--reaction time is shorter than the time during which the
(unperturbed) background field accesses the tachyonic regime. In this
non--perturbative regime, lattice simulation is necessary and
\verb|CosmoLattice| \cite{figueroaCosmoLattice2021} has been
used. Once the field fluctuations start to dominate the background,
the field fragments, see fig.~\ref{fig:0.01-fluc-spec}, and
non--linear soliton--like objects are formed, see
fig.~\ref{fig:0.01-snaps}. They are often called oscillons and can be
very long--lived, compared to the natural time scale of the dynamics
which is set by $1/m_\phi$. For our benchmark point
$\phi_0 = 0.01\ \mpl$ we computed the typical radius and energy
density of these oscillons, see fig.~\ref{fig:osci-M-R}; our results
qualitatively agree with similar results in the literature for
different potentials. It is worth noting that in contrast to a lot of
previous studies,
e.g. \cite{lozanovSelfresonanceInflationOscillons2018,
  lozanovEndInflationOscillons2014,
  antuschPropertiesOscillonsHilltop2019,
  aminFormationGravitationalClustering2019,
  zhouGravitationalWavesOscillon2013}, our inflaton potential is very
asymmetric around $\phi=0$; in particular, its second derivative is
negative only for positive $\phi \in ]\phi_0/3, \; \phi_0[$.

We saw that after the background field fragments, large fractions of
the total energy density are stored in the gradient and kinetic
energies. Nevertheless the system remains virialized throughout. The
equation of state parameter $\omega$ becomes slightly positive, see
fig.~\ref{fig:0.01-w-energy-frac}. The background evolution after the
collapse can be explained semi--quantitatively using the Hartree
approximation, see fig.~\ref{fig:0.01-hartree}. These features remain
qualitatively the same for yet smaller values of $\phi_0$, as shown in
sec.~\ref{sec:param}. However, the oscillon size and energy seem to
decrease somewhat with decreasing $\phi_0$, even when measured in
units of the appropriate power of $m_\phi$.

The inflaton fluctuations are highly non--Gaussian, see 
fig.~\ref{fig:0.01-delta-phi-2-3}. However, these perturbations are of
sub--horizon size after the end of inflation; the relevant length
scales are well below those probed by common cosmological observations
like the CMB anisotropies or even the Lyman$-\alpha$ forest. Moreover,
we expect that these inhomogeneities will be smoothed out once the
inflatons decay, which in our model happens well after the oscillons
decay by radiating inflaton particles.

Our lattice computation does not include gravitational effects.
Conceivably the large density fluctuations created when the inflaton
field fragments could seed primordial black hole formation.  We showed
that a single oscillon will not collapse into a black hole; however,
this does not exclude the possibility that the gravitational
attraction between oscillons, which was ignored in our simulation,
could eventually trigger gravitational collapse. We also noted that
black holes formed from a single, or a few, oscillons would decay well
before BBN via Hawking radiation. However, if most of the energy
density inside a Hubble radius at the end of the oscillon lifetime
should end up in a single black hole, it would be sufficiently heavy
to survive until today. While this scenario does not appear very
likely to us, non--perturbative computation including gravity
are likely necessary in order to settle this question.

In the turbulent phase when the background field fragments,
gravitational waves will be generated. They are (mostly) emitted
during oscillon formation, and are thus of sub--horizon size. As a
result, the predicted peak of the spectrum today lies in the MHz
range, which is very difficult to probe. Moreover, the energy density
of these gravitational has negligible impact on the expansion history
of the universe, see figs.~\ref{fig:gw-sens} and \ref{fig:gw_pl_fit}.

There are many avenues for possible future work. It might be possible
to improve the analytical understanding of the epoch of exponential
growth before back--reaction becomes important, via a more
sophisticated Floquet analysis. As already emphasized, we could not
make reliable prediction concerning the possible formation of
primordial black holes. To that end, computations in numerical
relativity will be required, see
e.g. \cite{nazariOscillonCollapseBlack2021,
  kouOscillonPreheatingFull2020}. Moreover, our numerical resources
did not allow us to extend the simulation beyond
${\cal O}(10^3) \ m_\phi^{-1}$. We expect interactions between
oscillons to continue at later times, possibly leading to clustering
and merging of oscillons
\cite{aminFormationGravitationalClustering2019,
  hahneScatteringCompactOscillons2020}. In fact, we do not even know
just how long our oscillons will survive even in the absence of
gravity.

In summary, we have shown that even in a renormalizable, small--field
model featuring a single real inflaton field the dynamics of the epoch
immediately following inflation can be very complicated. If the model
parameter $\phi_0 \lesssim 10^{-2} \ \mpl$, the inflaton field quickly
fragments into long--lived oscillons, leading to the production of
primordial gravitational waves at MHz frequencies, and perhaps even to
the formation of primordial black holes.

\appendix

\section*{Ackowledgement}
We would like to thank Yong Xu and Mustafa A. Amin for useful discussions. Our gratitude also goes to the {\tt CosmoLattice} team. We appreciate that the physics department of University Bonn provides us the computing cluster for the lattice simulation.

\section{Details of the Floquet analysis}
\label{sec:app-floquet}

Here, we try to describe the detailed computation of Floquet
exponents. This part is largely based on
\cite{aminNonperturbativeDynamicsReheating2015}.

\subsection{Finding the period}

In principle, one can find the period by computing the integral
\begin{equation}
  T = \sqrt{2} \int_{\phi_{\min}}^{\phi_{\max}} \frac {\dd{\phi}}
  {\sqrt{\rho_\tot - V(\phi)}}\,.
\end{equation}
If we take $\rho_\tot = V(\phi_0)$ and take $1 - \beta \approx 1$, we need to
evaluate the integral
\begin{equation} \label{math:floquet-period-ana}
  T = \sqrt{\frac{6}{d}} \frac{1}{\phi_0} \int_{x_{\min}}^{x_{\max}}
  \frac{\dd{x}}{\sqrt{1-3x^4 + 8x^3 - 6 x^2}}\,,
\end{equation}
where the new dimensionless variable $x = \phi/\phi_0$ has been
introduced. $x_{\max}$ and $x_{\min}$ are the maximal and minimal
field value in one oscillation. This integral does converge as long as
$x_\max < 1$. It is however, not so trivial to determine $x_{\min}$
and $x_{\max}$ analytically. For practical purposes, the period is
determined by solving the homogeneous differential equation for the
background field numerically.

\subsection{Solving the differential equations}

We write the original second order differential equation as two
coupled first--order equations; in matrix form:
\begin{equation} \label{math:x_eq}
\partial_t x(t) = U(t) x(t)\,,
\end{equation}
with
\begin{equation} \label{math:U_def}
  x =  \begin{pmatrix} \delta \phi_k \\ \delta \pi_k \end{pmatrix}; \quad
  U(t) = \begin{pmatrix} 0 & 1 \\ -k^2 - V''(\phi) & 0 \end{pmatrix} .
\end{equation}
The variable $\delta\pi_k = \delta \dot{\phi}_k$ is introduced for
convenience. 

Introduce the fundamental matrix such that\footnote{Note that off-diagonal
  elements of $\mathcal{O}$ can have non--zero mass dimensions!}
\begin{equation}
x(t) = \mathcal{O}(t,t_0) x(t_0)\,
\end{equation}
with boundary condition
\begin{equation} \label{math:O_init}
\mathcal{O}(t_0, t_0) = \id\,.
\end{equation}
From eq.(\ref{math:x_eq}) we have
\begin{equation} \label{math:O_diff}
\partial_t \mathcal{O}(t, t_0) = U(t) \mathcal{O}(t, t_0)\,.
\end{equation}

The Floquet theorem states that the fundamental matrix can be
expressed as
\begin{equation}
\mathcal{O}(t, t_0) = P(t, t_0) \exp[(t-t_0)\Lambda(t_0)]\,
\end{equation}
where $\Lambda(t_0)$ is defined via
$\mathcal{O}(t_0 + T, t_0) = \exp[T \Lambda(t_0)]$, and $P$ is
periodic, i.e. $P(t+T, t_0) = P(t, t_0)$. Both $P(t,t_0)$ and
$\Lambda(t_0)$ are matrices. The eigenvalues of $\Lambda(t_0)$ are the
Floquet exponents. $x(t)$ can then be expressed in terms of the
eigenbasis of $\Lambda(t_0)$:
\begin{align*}
	x(t) &= \mathcal{O}(t, t_0) x(t_0), \\
	  &= \sum_{s=1}^{2} P(t, t_0) c_s \exp[(t-t_0)\Lambda(t_0)] e_s(t_0), \\
	  &= \sum_{s=1}^{2} c_s P(t, t_0) e_s(t_0) e^{\mu_k^s (t-t_0)}\,,
\end{align*}
with $x(t_0) = \sum_{s=1}^{2} c_s e_s(t_0)$ and
$\Lambda(t_0) e_s (t_0) = \mu_k^s (t_0) e_s(t_0)$\footnote{One can see
  this from $Q^{-1} A Q = A_{\diag} \Rightarrow A Q = Q A_{\diag} $ and
  $Q$'s columns are the eigenvectors.}.

After diagonalizing the fundamental matrix
$\mathcal{O}(t_0 + T, t_0)$, the real part of the Floquet exponents
are
\begin{equation}
\realp(\mu_{k}^s) = \frac{1}{T} \ln |o_k^s|\,,
\end{equation}
with $o^s_k$ the eigenvalues of $\mathcal{O}(t_0 + T, t_0)$.

\subsection{Abel's identity}

As a consequence of Abel's identity, the sum of Floquet exponents must
be zero \cite{aminScaleDependentGrowthTransition2012}
\begin{equation}
  \sum_{s=1}^{2} \mu_k^s = \frac{1}{T} \ln (\prod_{s=1}^{2} o_k^s)
  = \frac{1}{T} \ln \det \mathcal{O}\,.
\end{equation}
Since the trace of $U$ in \eqref{math:U_def} vanishes, the Wronski
determinant computed from the $n$ eigenvalues of $\mathcal{O}$ must
by conserved by Abel's identity
\cite{teschlgeraldOrdinaryDifferentialEquations2020}:
\begin{equation}
W(t) = W(t_0) \exp(\int_{t_0}^t \dd{t'} \tr(U(t')))\,,
\end{equation}
where we have used eq.(\ref{math:O_diff}).
From the initial conditions \eqref{math:O_init}, we then have
\begin{equation}
\sum_{s=1}^{2} \mu_k^s = \frac{1}{T} \ln \det \id = 0\,.
\end{equation}
This can be used as a consistency check after computing each
individual Floquet exponent. The differential equations seem really
stiff; thus we choose to use the implicit Runge--Kutta method of order
$5$ implemented in \cite{2020SciPy-NMeth, harris2020array} for the
numerical solution.

\section{Lattice parameters}
\label{sec:app-lat-para}
\begin{table}[h!]
\centering
\begin{tabular}{c | c c c c c}
\toprule
$\phi_0 / \mpl$ & $0.1$ & $0.01$ & $5 \cdot 10^{-3}$ & $10^{-3}$ & $10^{-4}$ \\
\midrule
\verb|N| & $256$ & $256$ & $256$ & $256$ & $256$\\
\verb|tMax| & $200$ & $200$ & $300$ & $500$ & $500$\\
\verb|dt| & $0.005$ & $0.005$ & $0.005$ & $0.005$ & $0.005$\\
\verb|kIR| & $0.25$ & $0.25$ & $0.25 $ & $0.25 $ & $0.25 $\\
\verb|kCutoff| & $55$ & $55$ & $55$ & $55$ & $55$\\
$\verb|init. amp.| /\si{\giga\eV}$ & \num{1.151318366512767e+17} & \num{1.159073404226507e+16} & \num{5.737436015394929e+15} & \num{1.148838722103166e+15} & \num{1.113006825587231e+14}\\
$\verb|init. mom.| /\si{\giga\eV\tothe{2}}$ & \num{-7.071988905948155e+25} & \num{-7.410370315472198e+22} & \num{-9.271684888278322e+21} & \num{-7.422683904031192e+19} & \num{-7.667551290716926e+16}\\
\bottomrule
\end{tabular}
\caption{Lattice parameters for various $\phi_0$'s. Except for the
    initial amplitudes (\texttt{init. amp.}) and initial momenta (\texttt{init. mom.}), all quantities
are given in program units.}
\label{tab:phi0-0.01}
\end{table}

\section{Virial theorem}
\label{sec:app-virial}

Heuristically, one can derive the Virial theorem from the (full)
equation of motion
\begin{equation}
  \ddot{\phi} - \frac{1}{a^2} \nabla^2 \phi + 3H \dot{\phi} + \pdv{V}{\phi} = 0
  \,.
\end{equation}
Ignore the Hubble term for now. One can multiply all remaining terms
with $\phi$ and take the volume average:
\begin{equation}
  \expval{\phi \ddot{\phi}} - \frac{1}{a^2} \expval{\phi \nabla^2 \phi}
  + \expval{\phi \pdv{V}{\phi}} = 0\,.
\end{equation}
Assuming the surface terms will fall off fast enough, one obtains the
virial theorem after integration by parts and averaging over time:
\begin{equation}
  \frac{1}{2} \expval{\dot{\phi}^2} - \frac{1}{2a^2}
  \expval{\left(\nabla \phi \right)^2} -
  \frac{1}{2} \expval{\phi \pdv{V}{\phi}} = 0\,,
\end{equation}
where the three terms can be interpreted as kinetic, gradient and
potential energy, respectively.

This heuristic derivation neglects the Hubble friction term and the
surface terms in the integration by parts. More rigorously, one can
show the same relation by using the general virial theorem in
statistical physics
\begin{equation}
  \expval{\eta_i(x) \pdv{\mathcal{H}}{\eta_j(y)}} = T \delta_{ij} \delta(x - y)
  \,,
\end{equation}
with $\eta_i = \phi, \dot{\phi}$ \cite{schwabl, Boyanovsky_2004}. 

\section{Propagation of gravitational waves}
\label{sec:app-gw-prop}

In order to obtain the current frequency and scaled energy density of
GWs, one has to compute the redshifting due to the expansion of the
universe between the time of GW production and today. Note that
entropy is not conserved until thermalization by inflaton decay has
completed. We therefore factorize the relevant ratio of scale factors,
for later convenience we also introduce additional factors involving
densities that multiply to unity:
\begin{equation} \label{math:ap/a0}
  \frac{a_p}{a_0} = \frac{a_p}{a_{\text{th}}} \cdot \frac{a_{\text{th}}}{a_0}
  \cdot \frac{1}{\rho_{c, p}^{1/4}} \left( \frac{\rho_{c, p}}{\rho_{c, \text{th}}}
  \right) ^{1/4} \rho_{c, \text{th}}^{1/4}\,.
\end{equation}
Here $\rho_{c, p}$ and $\rho_{c, \text{th}}$ refer to the critical
energy density of the Universe at time of GW production and
thermalization, respectively; since to excellent approximation
$\Omega_{\rm tot} = 1$ after inflation, the critical energy density
can be equated with the total energy density.

For a general time-dependent equation of state parameter $\omega(t)$,
one has
\begin{align}
  \frac{\rho_{p}}{\rho_{\text{th}}} =
  \exp[-3\int^{a_p}_{a_{\text{th}}}(1+\omega) \dd(\ln \tilde{a} )]
  = \left( \frac{a_p} {a_{\rm th}} \right)^{-3(1+w)}\, .
\end{align}
In the second step we have made the
simplifying assumption that $\omega$ is constant till thermalization.
This yields
\begin{equation} \label{math:ap-factor1}
  \frac{a_p}{a_{\text{th}}} \left( \frac{\rho_{c, p}}{\rho_{c, \text{th}}}
  \right)^{1/4} = \left( \frac{a_p}{a_{\text{th}}} \right)^{(1-3\omega)/4}
  \lesssim 1 \,.
\end{equation}
Standard thermalization takes the Universe from matter domination
(coherent, harmonic oscillation of the inflaton field) to radiation
domination (the hot Big Bang), in which case the average value of
$\omega$ should be somewhere in between $0$ and $1/3$. In our case the
initial oscillations are quite anharmonic, leading to $\omega < 0$
initially, see fig.~\ref{fig:0.01-w-energy-frac}; later $\omega$ is
positive but quite small. For the sake of giving a definitive estimate
of the current peak frequency of GW, we take this factor to be unity.

If the entropy is conserved, then
\begin{equation}
S \propto g_{*, s}(T) T^3 a^3 = \const\,.
\end{equation}
As long as $g_{*, s} \simeq g_{*}$ the (relativistic) energy
density\footnote{Only radiation contributes to entropy at leading
  order.} scales like
\begin{equation}
\rho \propto g_{*} T^4 \propto g_*^{-1/3} a^{-4}\,.
\end{equation}
Since at reheating, $\rho_{\rm rel,th} \simeq \rho_{\rm tot,th} =
\rho_{c,{\rm th}}$, we have
\begin{equation} \label{math:ap-factor2}
  \frac{a_{\text{th}}}{a_0} = \left( \frac{\rho_{\text{rel}, 0}}
    {\rho_{c, \text{th}}} \right)^{1/4} \left(
    \frac{g_{*, 0}}{g_{*, \text{th}}} \right)^{1/12}\,.
\end{equation}
Inserting eqs.\eqref{math:ap-factor1} and \eqref{math:ap-factor2} back
into \eqref{math:ap/a0}, one has
\begin{equation} \label{math:rat_p_0}
  \frac{a_p}{a_0} \simeq \left( \frac{\rho_{\text{rel}, 0}}{\rho_{c, p}}
  \right)^{1/4} \left( \frac{g_{*, 0}}{g_{*, \text{th}}} \right)^{1/12}\,.
\end{equation}
The last factor depends on the reheat temperature only very weakly,
due to the power $1/12$. Even if all Standard Model degrees of freedom
contribute at $T_{\rm th}$,
$g_{*, 0} / g_{*, \text{th}} \approx 0.032$ in which case the last
factor in eq.(\ref{math:rat_p_0}) is about $0.75$; for $T_{\rm th}$
below the mass of the top quark this factor would be even closer to
$1$.

Today's radiation energy density can be computed from the CMB monopole
temperature
\begin{equation}
  \rho_{\text{rel}, 0} = 2 \frac{\pi^2}{30} T_0^4
  = \SI{1.928e-15}{\eV\tothe{4}}\,.
\end{equation}
Hence,
\begin{equation}
  \frac{a_p}{a_0} \simeq \num{1.571e-13} \frac{\si{\giga \eV}}{\rho_{c, p}^{1/4}}
  \,.
\end{equation}

The possibility to detect GWs crucially depends on their frequency
today. For a metric perturbation of fixed comoving size $k_\gw$
today's frequency is
\cite{lozanovGravitationalPerturbationsOscillons2019}
\begin{equation} \label{math:app-f0-og}
  f_0 = \frac{1}{2\pi} \frac{k_\gw}{a_p} \frac{a_p} {a_0}
  \simeq \frac{k_\gw}{a_p} \frac{1}{\rho_{c, p}^{1/4}} \cdot \num{3.8e10}\
  \si{\hertz}\, .
\end{equation}
The first factor is just the physical wavelength of the GW at time of
production. The physical interpretation of the $\rho_{c, p}^{-1/4}$
factor is clear: the higher the energy scale at production, the more
expansion the mode would experience in the past, leading to a longer
wavelength and lower frequency. Note that $\rho_{c, p}$ in the
numerical result is taken at the same time as the GW
spectrum.

Of course, only sufficiently strong GW signals are detectable. GWs are
radiation, thus their energy density decays like $a^{-4}$. At a
general temperature after production, the scaled energy density
becomes
\begin{align} \label{math:gw-Omega-T}
  \Omega_{\gw} (T) h^2 &= \Omega_{\gw, p} h^2 \frac{\rho_{c, p}}{\rho_c (T)}
                         \left( \frac{a_p}{a (T)} \right) ^4, \notag \\
  &= \Omega_{\gw, p} h^2 \frac{\rho_{\text{rel}}(T)}{\rho_{c}(T)}
  \left( \frac{g_{*}(T)}{g_{*, \text{SM}} } \right)^{1/3}\,,
\end{align}
where eq.\eqref{math:ap-factor2} has been used in the second step. Under the
above assumptions, this yields for the current Universe
\begin{equation} \label{math:gw-Omega-0}
\Omega_{\gw, 0} h^2 = \num{7.808e-6} \cdot \Omega_{\gw, p} h^2\, .
\end{equation}

\section{Dependence on the reheating temperature}
\label{sec:app-gw-reheating}

So far we have assumed that the factor appearing in
eq.\eqref{math:ap-factor1} can be set to unity. This allowed us
to derive expressions for today's peak frequency and energy density of GWs
that are independent of the reheat temperature. In this Appendix we
estimate this factor slightly more carefully.

\begin{figure}[ht]
\centering
\includegraphics[width=0.7\textwidth]{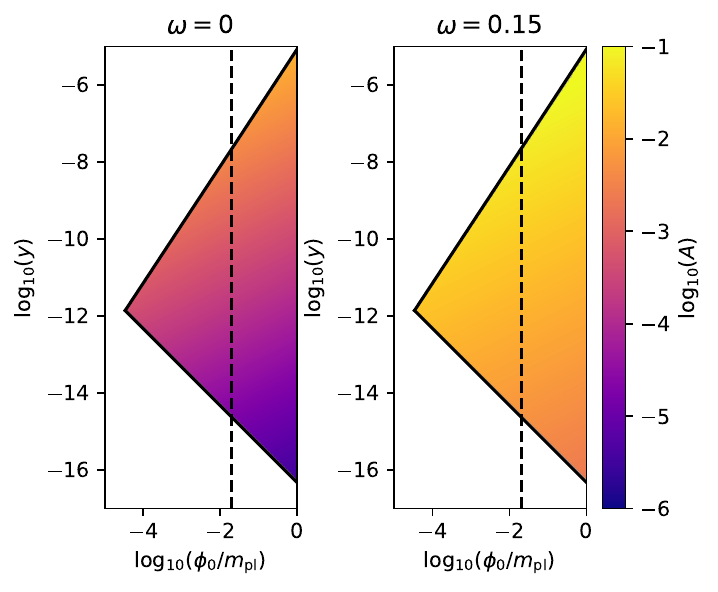}
\caption{Results for the factor $A$ defined in
  eq.(\ref{math:gw_extra_A}) for constant $\omega = 0$ (left) and
  $\omega = 0.15$ (right) from the time of GW production until the end
  of reheating. The latter proceeds through perturbative inflaton
  decay via a Yukawa coupling $y$. The upper and lower bounds on $y$
  come from the radiative stability of the inflaton potential and
  successful Big Bang nucleosynthesis, respectively
  \cite{dreesSmallFieldPolynomial2021}. The color corresponds to the
  logarithm of $A$. GW production from the formation of oscillons is
  significant only to the left of the dashed lines..}
\label{fig:gw-factor-A}
\end{figure}

For a general $\omega$ parameter, we have
\begin{equation}
  \frac{a(t)}{a_i} = \left[ \frac{3(\omega + 1)}{2} H_i (t - t_i)
    + 1\right] ^{\frac{2}{3(\omega+1)}}\,,
\end{equation}
where $t_i$, $a_i$ and $H_i$ are the initial cosmic time, the initial
scale factor and Hubble parameter, respectively. As a generous
estimate, we take $H_\gw \simeq H_I$ and
$t_{\text{reheating}} = 1 /\Gamma_\phi = 8\pi / y^2 m_\phi$, where we
have computed the inflaton decay width $\Gamma_\phi$ under the
assumption that it decays into a fermion pair via a Yukawa coupling
$y$ \cite{dreesSmallFieldPolynomial2021}. Thermalization is assumed to
be completed instantaneously after perturbative reheating. Then we define 
\begin{equation} \label{math:gw_extra_A}
  A := \left( \frac{a_p}{a_{\text{th}}} \right) ^{(1-3\omega)/4} \simeq
  \left[ \frac{3(\omega+1)}{2} \frac{\sqrt{d} \phi_0^2}{3 \mpl}
    \frac{8\pi}{y^2 \cdot 2 \sqrt{d} \phi_0 }
  \right]^{\frac{-1+3\omega}{6(1+\omega)}}
  = \left[ 2\pi (1+\omega) \frac{\phi_0}{\mpl} \frac{1}{y^2}
  \right]^{\frac{-1+3\omega}{6(1+\omega)}}.
\end{equation}
Evidently, if we had radiation domination, $\omega = 1/3$, during
perturbative reheating, this factor would be unity, since then
$\rho_\gw$ would redshift exactly like $\rho_{\rm tot}$. For matter
domination, $\omega=0$, the exponent is $-1/6$.  Putting in this
number, the left frame of fig.~\ref{fig:gw-factor-A} shows that $A$
can be up to five orders of magnitudes lower than unity. Thus, the GW
frequency will get shifted to smaller values, where planned
experiments are more sensitive; however, $\Omega_\gw$ will be reduced
as well. As indicated in eq.(\ref{math:gw_extra_A}), $A$ becomes
larger with decreasing $\phi_0$, which decreases the expansion rate
during this epoch, but decreases with decreasing $y$, which delays the
completion of reheating.

Recall that $\omega$ is slightly positive while the oscillons exist,
see fig.~\ref{fig:0.01-w-energy-frac}; moreover, more realistically a
significant radiation component will build up before reheating is
completed. Both effects will tend to increase $\omega$, and thus
$A$. An example is shown in the right frame of
fig.~\ref{fig:gw-factor-A}, where we have set $\omega = 0.15$. This
changes the power in eq.\eqref{math:gw_extra_A} to about $-0.08$,
leading to a much weaker dependence on $\phi_0$ and $y$ than in the
case with $\omega = 0$. Now $A$ is typically ${\cal O}(0.1)$ or
slightly smaller.

The exact value of $\omega$ is thus quite important here. As already
noted, during oscillon formation $\omega$ is positive but small.
Oscillons will eventually decay. According to
\cite{hertzbergQuantumRadiationOscillons2010,
  salmiRadiationRelaxationOscillons2012a,
  zhangClassicalDecayRates2020,
  lozanovEnhancedGravitationalWaves2022}, this happens via the
emission of inflaton particles with typical momentum $p \sim m_\phi$;
this should lead to an at least temporary increase of $\omega$.
Finally, as already noted, a precise computation of $A$ would have to
include the increase of $\omega$ due to the gradual build--up of the
radiation component.

\printbibliography
\end{document}